\documentclass[useAMS,usegraphicx,referee]{mn2e}

\title[AGN and starburst in ULIRGs]{Exploring the active galactic nucleus and starburst content of local ultraluminous infrared galaxies through 5--8~$\bmu$m spectroscopy}
\author[E. Nardini et al.]{E.~Nardini,$^1$\thanks{E-mail: nardini@arcetri.astro.it}
G.~Risaliti,$^{2,3}$ M.~Salvati,$^2$ E.~Sani,$^1$
\newauthor Y.~Watabe,$^2$ A.~Marconi$^1$ and R.~Maiolino$^4$\\
$^1$ Dipartimento di Astronomia e Scienza dello Spazio, Universit\`a di Firenze, L.go E. Fermi 2, 50125 Firenze, Italy\\
$^2$ INAF - Osservatorio Astrofisico di Arcetri, L.go E. Fermi 5, 50125 Firenze, Italy\\
$^3$ Harvard-Smithsonian Center for Astrophysics, 60 Garden St. Cambridge, MA 02138 USA\\
$^4$ INAF - Osservatorio Astronomico di Roma, via di Frascati 33, 00040 Monte Porzio Catone (RM), Italy}

\begin{document}

\date{Released 2009 Xxxxx XX}

\pagerange{\pageref{firstpage}--\pageref{lastpage}} \pubyear{2009}

\maketitle

\label{firstpage}

\begin{abstract}
We present a 5--8~$\mu$m analysis of the \textit{Spitzer}-IRS spectra of 71 
ultraluminous infrared galaxies (ULIRGs) with redshift $z<0.15$, devoted to the 
study of the role of active galactic nuclei (AGN) and starbursts (SB) as the 
power source of the extreme infrared emission. Around $\sim$5~$\mu$m an AGN 
is much brighter (by a factor $\approx$30) than a starburst of equal bolometric luminosity. 
This allows us to detect the presence of even faint accretion-driven cores inside ULIRGs: 
signatures of AGN activity are found in $\sim$70 per cent of our sample (50/71 sources). 
Through a simple analytical model we are also able to obtain a \textit{quantitative} 
estimate of the AGN/SB contribution to the overall energy output of each source. 
Although the main fraction of ULIRG luminosity is confirmed to arise from star 
formation events, the AGN contribution is non-negligible ($\sim$23 per cent) and 
is shown to increase with luminosity. The existence of a rather heterogeneous 
pattern in the composition and geometrical structure of the dust among ULIRGs 
is newly supported by the comparison between individual absorption features 
and continuum extinction. 
\end{abstract}


\section{Introduction}

Since a significant population of ultraluminous infrared galaxies (ULIRGs, i.e. 
the sources whose 8--1000~$\mu$m luminosity exceeds $L_\mathit{IR}>10^{12} L_\odot$) 
was discovered during the IRAS all-sky survey (Houck et al. 1985), the origin 
of their huge infrared (IR) emission has been widely debated. Extreme IR activity 
is known to be closely 
related to interacting or merging systems: optical and near-IR imaging shows that 
a large part of ULIRGs is indeed resulting from the collision of 
gas-rich galaxies (e.g. Clements et al. 1996; Scoville et al. 2000). 
Multiwavelength observations have pointed out that the typical spectral energy 
distribution (SED) of ULIRGs is characterized by a prominent far-IR peak due to 
dust reprocessing of higher-frequency primary radiation, whose elusive source 
is supposed to be a combination of merger-induced bursts of star formation and 
gas accretion onto a supermassive black hole. Despite their low frequency and 
limited contribution to the energy density of the local Universe 
(Soifer \& Neugebauer 1991), nearby ULIRGs represent the counterparts of both 
the submillimetre sources (Blain et al. 2002, and references therein) and the 
24~$\mu$m--selected galaxies that are responsible for the dominant energy output 
at high redshift (Caputi et al. 2007). Understanding their nature and physical properties 
is therefore a task of great importance, with far-reaching implications for models 
of structure formation and galaxy evolution. In particular, many efforts have been 
made in order to assess the contribution of the starburst (SB) and active galactic 
nucleus (AGN) components to ULIRGs: such an estimate would give deeper insight into 
the history of star formation over cosmic time and the origin of the X-ray and 
far-IR backgrounds. \\
All the diagnostic studies concerning ULIRGs have to come to terms with the great opacity 
of their nuclear regions, that precludes a straight identification of the hidden power 
source. Nevertheless the mid-IR thermal emission is expected to suffer from little 
extinction, and at the same time to preserve some relic signatures of the underlying 
engine, making this wavelength range very attractive for solid ULIRG diagnostics. 
With the coming of the \textit{Spitzer Space Telescope} (Werner et al. 2004) and its 
Infrared Spectrograph (IRS; Houck et al. 2004), high-quality mid-IR spectra of several 
tens of local ULIRGs have become available. We have therefore been able to perform a 
5--8~$\mu$m spectral analysis of a sample of 71 local ULIRGs, separating the AGN and SB 
components in the observed emission and providing a quantitative estimate of the AGN/SB 
contribution to the bolometric luminosity of each source\footnote{The bolometric 
luminosity of a ULIRG is usually dominated by its 8--1000~$\mu$m emission: throughout 
this paper the terms \textit{bolometric luminosity} and \textit{IR luminosity} are 
consequently used with an equivalent meaning.}. The preliminary results have been 
anticipated in a recent letter (Nardini et al. 2008, hereafter Paper~I), mainly 
focused on the presentation of the decomposition method. In the present work we will 
discuss in detail our findings and investigate their implications. The outline of 
this paper is the following: in Section~\ref{rd} we briefly review the main studies, 
from radio to X-rays, that have contributed to clarify the multiwavelength picture 
of ULIRGs. Section~\ref{sd} illustrates the physical motivations for 5--8~$\mu$m 
diagnostics and describes how recent \textit{Spitzer} observations have prompted 
our AGN/SB decomposition effort. The ULIRG sample is presented in Section~4, while 
Section~5 concerns observations and data reduction. In Section~6 we summarize the 
fitting model and all the analytical steps leading to the estimate of the AGN/SB 
contribution to the bolometric luminosity. The comparison between our results and 
those obtained at different wavelengths is provided in Section~7, and the main 
implications of our findings are discussed in Section~8. The conclusions are drawn 
up in Section~9, and all the fits are shown in the Appendix along with further 
considerations on some notable objects.

\section{Review of ULIRG diagnostics}
\label{rd}

The detection of a faint or obscured AGN and the estimate of its contribution to 
the bolometric luminosity of the hosting galaxy are the main open issues in the study 
of ULIRGs. This task is complicated because of the dust reprocessing, that smooths the 
intrinsic differences between the primary AGN and SB radiation fields. A detailed knowledge 
of the broadband intrinsic emission of both active galactic nuclei and starbursts is then 
essential in order to find some signatures of the buried energy source. \\
In the optical ULIRGs are classified on the basis of intensity ratios, involving 
emission lines with different excitation energy to test the hardness of the radiation field. 
However, dust extinction and reddening affect the use and reliability of such optical 
diagnostics in various ULIRGs; moreover the possibility of a differential extinction of 
the AGN and SB components is difficult to take into account. Qualitatively, compelling 
evidence of AGN activity is found in a modest fraction of the 118 ULIRGs included in 
the IRAS 1~Jy sample (Kim \& Sanders 1998), with a possible trend of increasing AGN 
detection rate in sources of higher luminosity (Veilleux, Kim \& Sanders 1999). 
At $L_\mathit{IR}>2\times 10^{12} L_\odot$ nearly a half of IR galaxies exhibit the 
typical features of Seyfert galaxies, and the fraction of AGN-like objects is even 
larger if one includes the sources classified as Low-Ionization Nuclear Emission-line 
Regions (LINERs). The connection between AGN and LINERs is anyway controversial, 
because the ionization degree of the latter can be a possible consequence of the 
supernova-driven shocks and the violent galactic winds pervading a 
starburst as well (Taniguchi et al. 1999; Lutz, Veilleux \& Genzel 1999). This 
leaves unsolved the issue about the fraction of ULIRGs actually hosting an AGN. \\
The dust opacity affecting the optical, ultraviolet (UV) and near-IR domains 
turns irrelevant at radio wavelengths, which in principle also benefit of a 
much higher angular resolution accessible through interferometry. Unfortunately 
these advantages are reduced by the low radio emission of ULIRGs. Nevertheless, 
although jet-like structures or extended lobes have never been found in ULIRGs, 
many of them present compact radio cores whose power does not correlate with 
the H$\alpha$ luminosity nor with other usual indicators of star formation 
(Nagar et al. 2003). Better spatial resolution is required to confirm the AGN 
nature of such sources and rule out any hypothesis of a compact cluster of stars. \\
Speculatively, the optimal spectral region to clear all the ambiguities and detect 
the AGN component is the 2--10~keV band, where it is possible to look through the 
obscuring material and take advantage of the sharp difference between the hard X-ray 
features and luminosity of starbursts and AGN. As a matter of fact, the circumnuclear 
mixture of gas and dust can be Compton-thick, with equivalent column densities in excess 
of $10^{24}$~cm$^{-2}$, and bury an AGN even in this energy range. Moreover ULIRGs are 
faint emitters, since their typical hard X-ray flux is usually far less than $\sim$10$^{-3}$ 
times their IR flux (e.g. Risaliti et al. 2000). Hence the current instrumentation 
allows the X-ray diagnostics 
for a handful of bright sources only: intense star formation activity is confirmed to 
be the dominant power supply for extreme IR luminosity, and even if convincing traces 
of elusive AGN (i.e. missed in the optical, Maiolino et al. 2003) are found, the 
conclusions are still tentative (Franceschini et al. 2003; Ptak et al. 2003). \\
In the mid- and far-IR, the reprocessed radiation still provides some clues to 
the enigmatic energy source, that can be recognized e.g. in the circumnuclear 
dust composition, in the SED large-scale profile and/or in the appearance of 
coronal lines. An empirical indicator of star formation is usually considered 
to be the group of emission features centred at 3.3, 6.2, 7.7, 8.6 and 11.3~$\mu$m, 
and attributed to polycyclic aromatic hydrocarbons (PAHs): these features are prominent 
in dusty environments exposed to soft UV radiation but strongly suppressed in the 
presence of harder radiation fields, probably because of the alteration/destruction 
of their carriers (Voit 1992). The concurrent use of PAH strength and appropriate 
line ratios, e.g. [O~\textsc{iv}]/[Ne~\textsc{ii}] and [Ne~\textsc{v}]/[Ne~\textsc{ii}], 
results in an effective probe into the nature of ULIRGs (Genzel et al. 1998). 
Yet the diagnostic methods based exclusively on emission lines can not be easily 
applied to faint sources. This involves the necessity of introducing the continuum 
properties and mid-IR colours as alternative diagnostic tools, as suggested by Laurent 
et al. (2000), with the replacement of the line ratio with the ratio between the warm 
and the hot dust continua. A critical shortcoming actually remains, consisting in the 
possible non-detection of a weak AGN inside the ULIRG: an unresolved AGN component which 
is heavily absorbed or intrinsically feeble can vanish within the much more diffuse and 
intense starburst emission.

\section{The AGN/SB spectral decomposition at 5--8~$\bmu$m}
\label{sd}

\begin{figure}
\includegraphics[width=8.5cm]{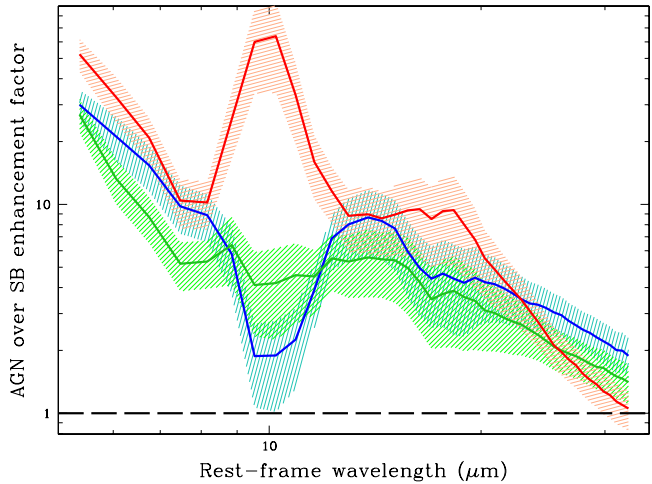}
\caption{Dependence from wavelength in the \textit{Spitzer}-IRS operative window of the AGN 
over SB brightness ratio for equal bolometric energy output. The curves are obtained by 
dividing the observed emission of IRAS~07598+6508 (\textit{red line}), IRAS~08572+3915 
(\textit{blue line}) and IRAS~12127$-$1412 (\textit{green line}) by that of our SB template, 
whose definition is described later on in this Section. The shaded regions show the 
1$\sigma$~\textit{rms} uncertainty, and the spectral resolution has been degraded in order 
to smooth the dependence of the ratios from narrow features. The three objects appear as 
AGN-dominated at a visual inspection, with moderate reddening, significant absorption and/or 
suppression of aromatic features. After our analysis, the AGN bolometric contribution to 
each source turns out to exceed 80 per cent. It is clear how the AGN spectral enhancement 
rapidly decreases with increasing wavelengths, giving rise to a dilution effect that 
precludes the detection of intrinsically faint or heavily obscured AGN components. Concerning 
the \textit{red curve}, the prominent 10~$\mu$m bump and the following plateau are exclusively 
due to the lack of silicate absorption, which is unusual in composite ULIRGs.} 
\label{enh}
\end{figure}

In order to overcome the potential AGN-dilution effect, it is necessary to select the 
wavelength range in which the imbalance between the AGN and SB components is maximum. By 
examining the average SEDs of both AGN and starbursts one immediately evinces that an 
AGN is much brighter than a starburst with equal bolometric luminosity not only in the 
hard X-ray domain, but also in a narrow IR band stretching from $\sim$3~$\mu$m, where the 
red tail of direct stellar emission begins to vanish, up to $\sim$8~$\mu$m, at the onset 
of the possible silicate absorption feature centred at 9.7~$\mu$m (see Fig.\ref{enh}). 
This difference can be ascribed to the more intense UV radiation field of the AGN, 
which is able to heat up to the sublimation temperature the dust located in proximity of 
the central energy source ($T_\mathit{sub} \sim$1500~K for graphite grains). Incidentally, 
both the above-mentioned mid-IR \textit{colour} diagnostic and the previous identification 
of a subset of \textit{warm} ULIRGs (Sanders et al. 1988) are based on a completely similar 
argument. 
It is only in the recent years that this spectral region and its potential in the study of 
ULIRGs have been exploited quantitatively (e.g. Risaliti et al. 2003, 2006a). 
In particular, \textit{L}-band spectroscopy ($\sim$3--4~$\mu$m) has proven to be a powerful method 
to uncover a buried AGN inside a ULIRG, thanks to diagnostic features like the equivalent 
width of the 3.3~$\mu$m PAH emission and the continuum slope (Risaliti et al. 2006b). 
Specifically, a steeply rising continuum hints at the reddening of a point-like energy 
source; an obscured AGN interpretation can be also supported by the simultaneous detection 
of deep absorption profiles at $\sim$3.4 and $\sim$4.5~$\mu$m due to aliphatic hydrocarbons 
and carbon monoxide, respectively (Sani et al. 2008). Ground-based observations are anyway 
affected by the strong variability of atmospheric transmission and the high thermal background. 
For this reason the quality of the current \textit{L}- and \textit{M}-band spectra, even if obtained with 
large-aperture facilities like SUBARU and VLT, allows the determination of the AGN and SB 
components for only the $\sim$20 brightest ULIRGs. In this view, the spectral 
capabilities at 2.5--5~$\mu$m of the AKARI satellite look very promising (e.g. Oyabu et al. 2009). \\
A significant progress has been possible thanks to \textit{Spitzer}, whose spectrograph 
operates longward of 5~$\mu$m and provides the natural extension of our diagnostic study. 
The luminosity enhancement of the AGN over the SB component is expected to substantially 
fade and turn over with increasing wavelength (Fig.\ref{enh}), none the less at 5--8~$\mu$m 
the considerably higher accuracy achieved from space still overrides this effect and allows 
a precise determination of the AGN and starburst properties in a composite ULIRG. 
Before proceeding with the details of our diagnostic technique, it is worth exploring the 
possibility for a luminous but extremely obscured AGN to be missed even at 5--8~$\mu$m. 
We consider the test case of a source whose IR luminosity arises in equal parts from 
nuclear activity and star formation. A contribution of 20 per cent to the observed 
5--8~$\mu$m emission can be assumed as a reasonable detection limit for the AGN component, 
which intrinsically is expected to outshine the starburst counterpart by a factor 
$\sim$20--30 in this spectral range. Consequently, the AGN should undergo a flux 
attenuation of $\sim$2 orders of magnitude in order to fall below the detection threshold. 
In terms of a standard extinction law this corresponds to a screen extinction of $A_V \sim$200~mag 
($A_{6\mu m} /A_V \sim$0.025; Nishiyama et al. 2008, 2009). This scenario is possible but unlikely 
(see also Section~\ref{ae} and the discussion in Genzel et al. 1998). We are then confident 
that virtually all the luminous AGN components inside ULIRGs can be probed through 
5--8~$\mu$m \textit{Spitzer}-IRS spectra. \\
The unprecedented sensitivity of the mid-IR capabilities onboard \textit{Spitzer}, 
in fact, has made it possible to reconstruct with high precision the SEDs of large 
samples of luminous active nuclei and starburst galaxies, and a surprising spectral homogeneity 
has been found within the two separate classes. At the highest luminosities this 
is especially true in the 5--8~$\mu$m range, with nearly constant shapes of both the 
aromatic complex in starbursts (Brandl et al. 2006, hereafter B06) and the continuum 
in AGN (Netzer et al. 2007, hereafter N07). Larger differences come out at longer 
wavelengths, where the temperature and emissivity of the different dust components as 
well as radiative transfer need to be taken into account. Our strategy then aims 
at disentangling the AGN and SB contributions to the observed 5--8~$\mu$m emission 
of ULIRGs by means of spectral templates, whose success hinges upon the large difference 
between the average AGN/SB emission and the small dispersion within the AGN and SB 
spectral classes. In the following we explain in detail the physical reasons which 
justify the selection of the AGN and SB templates for this work. \\

\begin{figure}
\includegraphics[width=8.5cm]{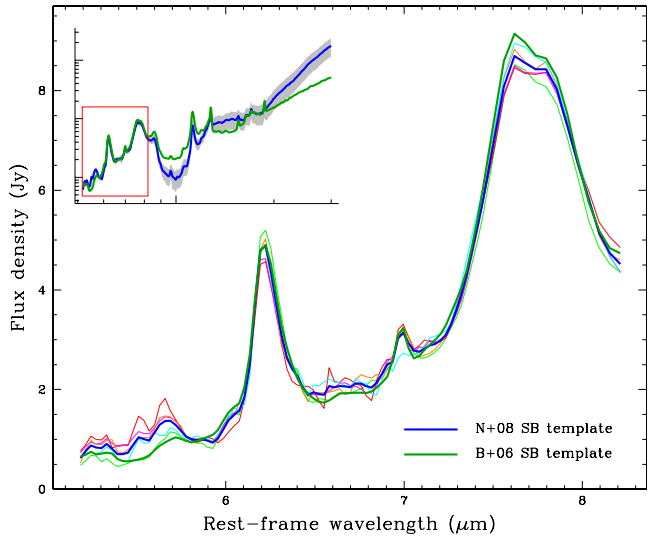}
\caption{Comparison between our SB template (\textit{blue solid line}) and that of B06 
(\textit{deep green dashed line}) at 5--8~$\mu$m. Their behaviour up to $\sim$30~$\mu$m 
is outlined in a log-log scale in the top left corner. The single spectra of the 
five sources employed in our template have also been plotted with lines of different 
colour: IRAS~10190$+$1322 (\textit{light green}), IRAS~12112$+$0305 (\textit{red}), 
IRAS~17208$-$0014 (\textit{magenta}), IRAS~20414$-$1651 (\textit{orange}) and 
IRAS~22491$-$1808 (\textit{cyan}).} 
\label{sbt}
\end{figure}

\textit{Starburst.} As previously mentioned, the mid-IR spectra of starburst  
galaxies display little variations from one object to another, especially in 
the 5--8~$\mu$m wavelength range. At first sight this appears to be a singular 
circumstance, because of the large number of physical variables involved in 
determining the observed properties of a starburst: these include the initial 
mass function, the duration and evolutionary stage of the single bursts, the 
properties of the dust grains and those of the interstellar environment, and 
obviously the total mass and luminosity of the system. It has already been 
suggested in B06 that such a remarkable similarity can derive from the spatial 
integration over a huge number of unresolved star-forming spots: this is 
indeed an interesting point that should be investigated further. 
The typical SB emission in the 5--8~$\mu$m interval is characterized by two 
prominent aromatic features centred at 6.2 and 7.7~$\mu$m. In SB-dominated 
ULIRGs the shape of these lines and their intensity ratio slightly differ 
from those observed in lower luminosity starbursts (like NGC~7714), probably 
because of higher obscuration in the former (Rigopoulou et al. 1999). Moreover 
the reliability of PAHs as quantitative tracers of star formation is still 
debated (Peeters, Spoon \& Tielens 2004; F\"orster Schreiber et al. 2004). 
In order to properly describe the SB component in our spectra, and check if the 
prescription of little dispersion may be applied in the ULIRG luminosity range as 
well, we performed a preliminary analysis on a representative subset of sources that 
other multiband diagnostics have established to host powerful starbursts. The 
5--8~$\mu$m emission of these sources has been approximated by means of two gaussian 
profiles for the aromatic features and a power-law continuum\footnote{Throughout 
this paper we describe the emission in terms of the flux density $f_\nu(\lambda)$, 
hence all the spectral indices of the power-law components in the following are 
defined as $f_\nu(\lambda) \propto \lambda^\Gamma$.}. As a result, 
the main properties of the PAH features turn out to be the same observed in lower 
luminosity systems. Their equivalent width is confirmed not to depend on the 
global IR output, and the modest spread concerning their relative strength has 
already been related to internal obscuration; the slope of the underlying 
continuum is $\Gamma_\mathit{sb} \ga$4, roughly uniform and well above the 
value assumed for the AGN component (see later). This tight correlation 
between the properties of the PAH features and the dust continuum is brought 
out also in B06, hence we are justified in the use of a spectral template 
to represent through a single scale factor the SB component of ULIRGs. 
A complete characterization of the whole SB spectrum would of course require a 
detailed model involving many parameters, and this indeed remains unavoidable 
when considering larger spectral domains. \\
Instead of adopting a prototypal SB template from the literature, we have 
constructed a new one as the average spectrum of the five brightest SB-dominated 
sources in our sample, i.e. IRAS~10190$+$1322, IRAS~12112$+$0305, 
IRAS~17208$-$0014, IRAS~20414$-$1651 and IRAS~22491$-$1808. The resulting 
slope is $\Gamma_\mathit{sb} \simeq$4.2, and the equivalent widths of the aromatic 
features are 0.43 and 0.88~$\mu$m: such values do not actually differ from the measures 
of B06, as can be easily evinced from Fig.\ref{sbt}. 
The comparison between our SB template and that of B06 over the entire IRS spectral 
range has already been shown in Paper~I, in order to emphasize their divergence 
just longward of $\sim$8~$\mu$m. This is anyway displayed again in the box of 
Fig.\ref{sbt}, where we focus instead on their similarity over the wavelengths 
of interest. Such similarity (and the fact that the PAH equivalent widths are 
the largest among ULIRGs) rules out any possibility of significant AGN contamination 
in our SB template\footnote{We also note that in this case our AGN/SB spectral 
decomposition would have brought out the necessity of a \textit{negative} AGN 
contribution in order to optimally fit the shape of a genuine SB-dominated ULIRG.}. \\
Summarizing, the spectral slope of the observed 5--8~$\mu$m ULIRG 
continuum is very sensitive to the possible AGN contribution. The PAH features 
and their intensities play of course an important role in modelling the spectral 
shape, but the dispersion in their properties which is found in pure starbursts is 
actually too small to alter the determination of the continuum, mimicking the 
presence of a bolometrically significant AGN in a composite source. \\

\textit{Active galactic nucleus.} At the wavelengths of concern the AGN emission 
is dominated by the cooling of small dust grains that are transiently heated up to 
temperatures close to the sublimation limit. Such a process is expected to produce 
a nearly featureless power-law continuum, and yet aromatic features are found in several 
quasars, as confirmed by recent observations (Schweitzer et al. 2006; Lutz et al. 2007, 
2008). The inferred star formation rate can be very high, and raises the question on the 
extent of the stellar contamination in the observed spectra of quasars and the reliability 
of a pure AGN template. In N07 the far-IR emission of a large set of Palomar-Green (PG) quasars 
is ascribed to cold dust in extended regions, and then suspected of having a stellar origin; 
the contribution of star formation to be subtracted in order to determine the intrinsic 
SED of a type 1 AGN is obtained from the properties of 12 SB-dominated ULIRGs. 
The resulting average spectrum can be modelled by a single power law from 
$\sim$3~$\mu$m to the 9.7~$\mu$m silicate bump (which usually appears in emission 
whenever the source is not obscured). The most striking outcome is that the spectral 
dispersion at these wavelengths is very small if compared to the large differences 
found in the far-IR. It is then reasonable to adopt a power-law component to reproduce 
the 5--8~$\mu$m AGN flux density. The choice of the spectral index however is not 
immediate, and has to be briefly discussed. A coarse estimate, as deduced from the 
SEDs provided by N07, yields $\Gamma_\mathit{agn} \sim$0.7; a better value can be 
anyway achieved from our \textit{L}-band study. The reason is that we are dealing with an 
active nucleus component in a ULIRG environment, and the overall dust properties 
are expected to differ from those of the PG quasars in the N07 sample. 
\begin{figure}
\includegraphics[width=8.5cm]{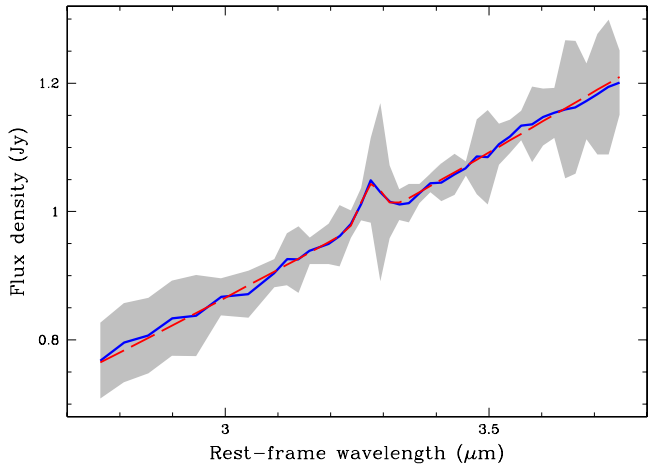}
\caption{The \textit{blue solid line} represents the average \textit{L}-band spectrum of a source 
whose near-IR emission is dominated by an unobscured AGN; the shaded area encompasses the 
1$\sigma$~\textit{rms} dispersion. The best fit (\textit{red dashed curve}) requires 
two components: a power law with spectral index 1.51$\pm$0.09, and a gaussian line with 
equivalent width $<3$~nm, far below the typical values of the PAH emission in starburst 
galaxies ($\sim$100~nm).}
\label{ind}
\end{figure}
In Fig.\ref{ind} we show the spectral template obtained by averaging the \textit{L}-band emission 
of four ULIRGs in our sample known to host a fairly unobscured active nucleus: MRK~231, 
IRAS~15462$+$0450 and IRAS~21219$-$1757, which are optically classified as type 1 Seyfert 
galaxies, and IRAS~05189$-$2524, which is a type 2. For all these sources the \textit{L}-band 
emission consists of a strong continuum arising from the dust heated by the AGN, as 
also confirmed by the small equivalent width of the 3.3~$\mu$m aromatic feature, only 
reaching a few nm while usually exceeding $\sim$100~nm in pure starbursts 
(Moorwood 1986; Imanishi \& Dudley 2000). The mean 
value of the continuum slope is $\sim$1.5, and such a value has then been assumed for 
the present analysis. Its extrapolation to 5--8~$\mu$m is supported by N07, since no 
significant change in the continuum gradient is expected. Moreover the same 5--8~$\mu$m 
spectra of the four template sources are fully consistent with the stretching of 
their \textit{L}-band slope, in spite of the growing aromatic emission. 
The slight discrepancy between these estimates of the AGN spectral index, based 
on different samples, seems to be a natural consequence of the greater obscuration 
characterizing the ULIRG environment, rather than of a significant stellar 
contamination even at such short wavelengths. As mentioned, the unobscured AGN components 
of this work and the PG quasars of N07 (even though intense star formation is in place) are 
different populations, in terms of both the luminosity of the AGN and the properties of the 
absorbing medium. A higher covering factor of the dust surrounding the AGN component in a 
ULIRG environment can account for a steeper intrinsic slope. The covering factor is indeed 
shown to decrease as a function of the AGN luminosity (Maiolino et al. 2007). \\ 
Beyond these aspects, another crucial effect has to be considered: the active nucleus 
component is still extremely compact, and the AGN-related hot dust emission can be 
extinguished by an external screen, consisting either of the outer layers of the 
obscuring torus predicted in unification models (Antonucci 1993) or of a thick absorbing 
cloud along the line of sight. We have therefore introduced a wavelength dependent 
attenuation factor $e^{-\tau(\lambda)}$, assuming for the optical depth a 
power-law behaviour $\tau(\lambda) \propto \lambda^{-1.75}$ (Draine 1989). 
On the contrary, the SB component can easily extend over several hundreds of 
pc and therefore can be affected exclusively by internal extinction, which is 
already embodied in the shape of the template: in fact, even if the SB 
extinction becomes apparent when considering the silicate absorption profiles 
at longer wavelengths, the 5--8~$\mu$m emission does not seem to be significantly 
modified except for the limited dispersion in the PAH ratio. \\
In Section~\ref{dis} we will deal again (from a quantitative point of view) with 
the intrinsic slope of the AGN continuum emission, investigating its interplay with the 
shape of the extinction law, and how any change can affect the AGN/SB spectral decomposition 
and the subsequent estimate of the relative AGN/SB bolometric contribution.

\section{The ULIRG sample}

\begin{table*}
\begin{center}
\caption{General properties of the 71 sources in our sample. $z$, $D_L$: Redshift and 
luminosity distance (in Mpc). $f_{12}$, $f_{25}$, $f_{60}$, $f_{100}$: IRAS flux densities 
(in Jy, with the colon indicating the revisions in Sanders et al. 2003). $L_\mathit{IR}$: 
Total IR luminosity computed according to Sanders \& Mirabel (1996). Class: Morphological 
classification as defined in Veilleux, Kim \& Sanders (2002). IIIa: Wide binary (apparent separation 
$>10$~kpc); IIIb: Close binary (apparent separation $<10$~kpc); IVa: Diffuse merger; 
IVb: Compact merger; V: Old merger. The entries within brackets are derived from Duc, 
Mirabel \& Maza (1997) and Scoville et al. (2000).}
\label{t1}
\begin{scriptsize}
\begin{tabular}{lccr@{.}lr@{.}lr@{.}lr@{.}lcc}
\hline \hline
\\
Object & $z$ & $D_L$ & \multicolumn{2}{c}{$f_{12}$} & \multicolumn{2}{c}{$f_{25}$} & 
\multicolumn{2}{c}{$f_{60}$} & \multicolumn{2}{c}{$f_{100}$} & log $(L_\mathit{IR}/L_\odot)$ & Class \\
\\
\hline
ARP 220 & 0.018 & 79 & 0 & 61: & 8 & 00: & 104 & 09: & 115 & 29: & 12.18 & IIIb \\
IRAS 00091$-$0738 & 0.118 & 554 & $<$0 & 07 & 0 & 22 & 2 & 63 & 2 & 52 & 12.27 & IIIb \\
IRAS 00188$-$0856 & 0.128 & 604 & $<$0 & 12 & 0 & 37 & 2 & 59 & 3 & 40 & 12.42 & V \\
IRAS 00456$-$2904 & 0.110 & 513 & $<$0 & 08 & 0 & 14 & 2 & 60 & 3 & 38 & 12.23 & IIIa \\
IRAS 00482$-$2721 & 0.129 & 608 & $<$0 & 10 & $<$0 & 18 & 1 & 13 & 1 & 84 & 12.09 & IIIb \\
IRAS 01003$-$2238 & 0.118 & 550 & $<$0 & 23 & 0 & 66 & 2 & 29 & 1 & 79 & 12.33 & V \\
IRAS 01166$-$0844 & 0.118 & 553 & $<$0 & 14 & 0 & 17 & 1 & 74 & 1 & 42 & 12.13 & IIIb \\
IRAS 01298$-$0744 & 0.136 & 644 & $<$0 & 12 & $<$0 & 28 & 2 & 47 & 2 & 08 & 12.37 & IVb \\
IRAS 01569$-$2939 & 0.141 & 669 & $<$0 & 11 & 0 & 14 & 1 & 73 & 1 & 51 & 12.27 & IVa \\
IRAS 02021$-$2103 & 0.116 & 541 & $<$0 & 07 & 0 & 30 & 1 & 45 & 1 & 72 & 12.09 & IVa \\
IRAS 02411+0353 & 0.144 & 683 & $<$0 & 08 & 0 & 22 & 1 & 37 & 1 & 95 & 12.28 & IIIb \\
IRAS 03250+1606 & 0.129 & 607 & $<$0 & 10 & $<$0 & 15 & 1 & 38 & 1 & 77 & 12.12 & IVb \\
IRAS 04103$-$2838 & 0.117 & 549 & 0 & 08 & 0 & 54 & 1 & 82 & 1 & 71 & 12.24 & IVb \\
IRAS 05189$-$2524 & 0.043 & 189 & 0 & 74: & 3 & 47: & 13 & 25: & 11 & 84: & 12.17 & IVb \\
IRAS 07598+6508 & 0.148 & 707 & 0 & 26 & 0 & 53 & 1 & 69 & 1 & 73 & 12.54 & IVb \\
IRAS 08559+1053 & 0.148 & 706 & $<$0 & 10 & 0 & 19 & 1 & 12 & 1 & 95 & 12.26 & IVb \\
IRAS 08572+3915 & 0.058 & 261 & 0 & 33: & 1 & 76: & 7 & 30: & 4 & 77: & 12.16 & IIIb \\
IRAS 09039+0503 & 0.125 & 587 & $<$0 & 17 & $<$0 & 21 & 1 & 48 & 2 & 06 & 12.17 & IVa \\
IRAS 09116+0334 & 0.145 & 691 & $<$0 & 09 & $<$0 & 14 & 1 & 09 & 1 & 82 & 12.18 & IIIa \\
IRAS 09539+0857 & 0.129 & 607 & $<$0 & 15 & $<$0 & 15 & 1 & 44 & 1 & 04 & 12.11 & V \\
IRAS 10190+1322 & 0.077 & 348 & $<$0 & 10 & 0 & 38 & 3 & 33 & 5 & 57 & 12.06 & IIIb \\
IRAS 10378+1109 & 0.136 & 645 & $<$0 & 11 & 0 & 24 & 2 & 28 & 1 & 82 & 12.36 & IVb \\
IRAS 10485$-$1447 & 0.133 & 628 & $<$0 & 11 & $<$0 & 30 & 1 & 73 & 1 & 66 & 12.23 & IIIa \\
IRAS 10494+4424 & 0.092 & 423 & $<$0 & 12 & 0 & 16 & 3 & 53 & 5 & 41 & 12.21 & IVb\\
IRAS 11095$-$0238 & 0.107 & 495 & $<$0 & 14 & 0 & 42 & 3 & 25 & 2 & 53 & 12.28 & IVb \\
IRAS 11130$-$2659 & 0.136 & 644 & $<$0 & 09 & 0 & 20 & 1 & 21 & 1 & 24 & 12.14 & IVa \\
IRAS 11387+4116 & 0.149 & 709 & $<$0 & 20 & $<$0 & 14 & 1 & 02 & 1 & 51 & 12.22 & V \\
IRAS 11506+1331 & 0.127 & 599 & $<$0 & 10 & $<$0 & 29 & 2 & 58 & 3 & 32 & 12.36 & IVb \\
IRAS 12072$-$0444 & 0.128 & 604 & $<$0 & 12 & 0 & 54 & 2 & 46 & 2 & 47 & 12.41 & IVb \\
IRAS 12112+0305 & 0.073 & 332 & $<$0 & 11: & 0 & 66: & 8 & 18: & 9 & 46: & 12.33 & IIIb \\
IRAS 12127$-$1412 & 0.133 & 628 & $<$0 & 13 & 0 & 24 & 1 & 54 & 1 & 13 & 12.20 & IIIa \\
IRAS 12359$-$0725 & 0.138 & 654 & $<$0 & 19 & $<$0 & 22 & 1 & 32 & 1 & 12 & 12.19 & IIIa \\
IRAS 13335$-$2612 & 0.125 & 587 & $<$0 & 13 & $<$0 & 14 & 1 & 40 & 2 & 10 & 12.13 & IIIb \\
IRAS 13454$-$2956 & 0.129 & 607 & $<$0 & 06 & $<$0 & 16 & 2 & 16 & 3 & 38 & 12.31 & IIIa \\
IRAS 13509+0442 & 0.136 & 643 & 0 & 10 & $<$0 & 23 & 1 & 56 & 2 & 23 & 12.30 & IVb \\
IRAS 13539+2920 & 0.108 & 504 & $<$0 & 09 & 0 & 12 & 1 & 83 & 2 & 73 & 12.09 & IIIb \\
IRAS 14060+2919 & 0.117 & 545 & $<$0 & 10 & 0 & 14 & 1 & 61 & 2 & 42 & 12.13 & IVa \\
IRAS 14197+0813 & 0.131 & 618 & $<$0 & 17 & $<$0 & 19 & 1 & 10 & 1 & 66 & 12.12 & V \\
IRAS 14252$-$1550 & 0.150 & 714 & $<$0 & 09 & $<$0 & 23 & 1 & 15 & 1 & 86 & 12.24 & IIIb \\
IRAS 14348$-$1447 & 0.083 & 377 & $<$0 & 10: & 0 & 55: & 6 & 82: & 7 & 31: & 12.36 & IIIb \\
IRAS 15130$-$1958 & 0.109 & 508 & $<$0 & 14 & 0 & 39 & 1 & 92 & 2 & 30 & 12.17 & IVb \\
IRAS 15206+3342 & 0.124 & 584 & 0 & 08 & 0 & 35 & 1 & 77 & 1 & 89 & 12.25 & IVb \\
IRAS 15225+2350 & 0.139 & 659 & $<$0 & 07 & 0 & 18 & 1 & 30 & 1 & 48 & 12.18 & IVa \\
IRAS 15250+3609 & 0.055 & 247 & 0 & 16: & 1 & 31: & 7 & 10: & 5 & 93: & 12.06 & (IIIb) \\
IRAS 15462$-$0450 & 0.100 & 461 & $<$0 & 13 & 0 & 45 & 2 & 92 & 3 & 00 & 12.22 & IVb \\
IRAS 16090$-$0139 & 0.134 & 631 & 0 & 09 & 0 & 26 & 3 & 61 & 4 & 87 & 12.57 & IVa \\
IRAS 16156+0146 & 0.132 & 623 & $<$0 & 10 & 0 & 28 & 1 & 13 & 1 & 00 & 12.12 & IIIb \\
IRAS 16468+5200 & 0.150 & 716 & $<$0 & 06 & 0 & 10 & 1 & 01 & 1 & 04 & 12.12 & IIIb \\
IRAS 16474+3430 & 0.111 & 519 & $<$0 & 13 & 0 & 20 & 2 & 27 & 2 & 88 & 12.21 & IIIb \\
IRAS 16487+5447 & 0.104 & 480 & $<$0 & 07 & 0 & 20 & 2 & 88 & 3 & 07 & 12.19 & IIIb \\
IRAS 17028+5817 & 0.106 & 492 & $<$0 & 06 & 0 & 10 & 2 & 43 & 3 & 91 & 12.18 & IIIa \\
IRAS 17044+6720 & 0.135 & 638 & $<$0 & 07 & 0 & 36 & 1 & 28 & 0 & 98 & 12.18 & IVb \\
IRAS 17179+5444 & 0.147 & 700 & $<$0 & 08 & 0 & 20 & 1 & 36 & 1 & 91 & 12.28 & IVb \\
IRAS 17208$-$0014 & 0.043 & 190 & 0 & 20: & 1 & 61: & 32 & 13: & 36 & 08: & 12.42 & (V) \\
IRAS 19254$-$7245 & 0.062 & 277 & 0 & 22 & 1 & 24 & 5 & 16: & 5 & 79 & 12.09 & (IIIb) \\
IRAS 20100$-$4156 & 0.130 & 610 & $<$0 & 13 & 0 & 34 & 5 & 19: & 5 & 16 & 12.65 & (IIIb) \\
IRAS 20414$-$1651 & 0.087 & 398 & $<$0 & 65 & 0 & 35 & 4 & 36 & 5 & 25 & 12.31 & IVb \\
IRAS 20551$-$4250 & 0.043 & 190 & 0 & 28: & 1 & 87: & 12 & 19: & 10 & 31: & 12.05 & (V) \\
IRAS 21208$-$0519 & 0.130 & 613 & $<$0 & 09 & $<$0 & 15 & 1 & 17 & 1 & 66 & 12.08 & IIIa \\
IRAS 21219$-$1757 & 0.112 & 521 & 0 & 21 & 0 & 45 & 1 & 07 & 1 & 18 & 12.14 & V \\
IRAS 21329$-$2346 & 0.125 & 587 & $<$0 & 08 & $<$0 & 16 & 1 & 65 & 2 & 22 & 12.16 & IVa \\
IRAS 22206$-$2715 & 0.131 & 620 & $<$0 & 10 & $<$0 & 16 & 1 & 75 & 2 & 33 & 12.23 & IIIb \\
IRAS 22491$-$1808 & 0.078 & 353 & $<$0 & 09 & 0 & 55 & 5 & 44 & 4 & 45 & 12.20 & IIIb \\
IRAS 23128$-$5919 & 0.045 & 198 & 0 & 35: & 1 & 64: & 10 & 94: & 10 & 68: & 12.06 & (IIIb) \\
IRAS 23234+0946 & 0.128 & 602 & $<$0 & 06 & $<$0 & 20 & 1 & 56 & 2 & 11 & 12.16 & IIIb \\
IRAS 23327+2913 & 0.107 & 496 & $<$0 & 06 & 0 & 22 & 2 & 10 & 2 & 81 & 12.13 & IIIa \\
MRK 231 & 0.042 & 187 & 1 & 83: & 8 & 84: & 30 & 80: & 29 & 74: & 12.55 & IVb \\
MRK 273 & 0.038 & 167 & 0 & 24: & 2 & 36: & 22 & 51: & 22 & 53: & 12.18 & IVb \\
NGC 6240 & 0.024 & 107 & 0 & 59: & 3 & 55: & 22 & 94: & 26 & 49: & 11.86 & (IIIb) \\
4C +12.50 & 0.122 & 570 & $<$0 & 14 & 0 & 67 & 1 & 92 & 2 & 06 & 12.32 & IIIb \\
UGC 5101 & 0.039 & 174 & 0 & 25: & 1 & 02: & 11 & 68: & 19 & 91: & 12.00 & (V) \\
\hline
\end{tabular}
\end{scriptsize}
\end{center}
\end{table*}

The present sample consists of 71 objects and is fully representative of the ULIRG 
population in the local Universe. In order to avoid a bias towards AGN activity, the  
flux limitation has been adopted at 60~$\mu$m. The 60~$\mu$m flux density is in fact 
a good proxy of the cold dust component, and this translates into a fairly unbiased 
selection with respect to the nature of the energy source. All the ULIRGs in the 
already mentioned 1~Jy sample with available \textit{Spitzer}-IRS observations and 
the additional requirement of a redshift $z<0.15$ have been selected, for a total of 
63 objects. With respect to Paper~I, three extra sources have been added, i.e. 
IRAS~02021$-$2103, IRAS~07598+6508 and IRAS~08559+1053. The broad absorption line 
(BAL) Seyfert~1 galaxy IRAS~07598$+$6508, that was initially dropped since a 
significant fraction of its mid-IR emission was expected to have a non-thermal origin, 
and IRAS~08559+1053, that was only observed during the in-flight calibration phase 
of \textit{Spitzer} (by J.R.Houck), have been retrieved for completeness. We point out 
that the volume restriction is absolutely arbitrary in this context, since it is 
directly imported from our previous studies, being it necessary to apply our \textit{L}-band 
diagnostics. The 1~Jy sample indeed contains 70 objects meeting this selection 
requirement. No observation is anyway available for the seven missing sources: only 
four has been recently observed with \textit{Spitzer}, but archive data are not 
provided as yet; the other three have not been observed. Within this uncovered subset 
five objects are optically classified as Seyfert~2 galaxies, while two are unclassified. 
The consequent bias is then minor, and will be discussed later on when we assess the 
incidence of black hole accretion and star formation at extreme IR luminosities. \\
Summarizing, all the selected targets obey the following criteria: \textit{(1)} a redshift 
$z<0.15$ and \textit{(2)} a 60~$\mu$m IRAS flux density $f_{60}>1$~Jy. The statistics is 
already large enough to derive general conclusions about the local ULIRG population. 
However, being the 1~Jy sample also characterized by constraints on the position of the target, 
a systematic search for ULIRGs inside the IRAS 2~Jy or 1.2~Jy all-sky samples (Strauss et al. 1992; 
Fisher et al. 1995, respectively) would yield a wealth of additional sources fulfilling 
the present criteria. We decided to complete our sample by adding only the eight 
sources drawn from the 2~Jy sample which are already present in Genzel et al. (1998), since 
they are among the brightest and best-studied ULIRGs. \\
The main properties of all the sources in our sample are listed in Tab.\ref{t1}. NGC~6240 
is the only source that does not strictly comply with the ULIRG classification, because of 
its slightly lower IR luminosity. It is anyway an advanced merger showing all the 
morphological and physical properties of ULIRGs, and is usually included in this class. 
Concerning this, it is also important to specify that the total IR luminosity of our 
sources has been computed according to the broadband flux equation in 
Sanders \& Mirabel (1996): 
\begin{equation}
F_\mathit{IR}=1.8 \times 10^{-11}(13.48 f_{12}+5.16 f_{25}+2.58 f_{60}+ f_{100}),
\label{e1}
\end{equation}
where $F_\mathit{IR}$ is the total IR flux in units of ergs s$^{-1}$ cm$^{-2}$ and 
$f_{12}$, $f_{25}$, $f_{60}$ and $f_{100}$ are the IRAS flux densities in Jy. When 
only an upper limit is available for $f_{12}$ (and $f_{25}$) we have simply assumed 
the one half of this limit. Such strategy provides a good approximation of 
the true flux density if compared to the few estimates proposed in Kim \& Sanders (1998). 
The luminosity distances have been obtained making use of a standard cosmology with 
$H_0=70$~km~s$^{-1}$~Mpc$^{-1}$, $\Omega_m=0.27$ and $\Omega_\Lambda=0.73$ 
(Hinshaw et al. 2009).

\section{Observations and Data Reduction}

\begin{table*}
\begin{center}
\caption{Observation log of our ULIRG sample. All the sources except one have been 
observed within the following programs: Spectroscopic study of distant ULIRGs II 
(PID 105, PI J.R.Houck), Buried AGN in ultraluminous infrared galaxies (PID 2306, 
PI M.Imanishi), The evolution of activity in massive gas-rich mergers (PID 3187, 
PI S.Veilleux). T$_{SL}$: Integration time in number of cycles times seconds for 
the IRS Short-Low orders (SL2: $\sim$5.2--7.7~$\mu$m; SL1: $\sim$7.4--14.5~$\mu$m). 
Since each cycle consists of two nod positions, the total observing time per slit is 
obtained by multiplying these entries by an additional factor 2.}
\label{t2}
\begin{scriptsize}
\begin{tabular}{lccclccc}
\hline \hline
\\
Object & PID & Date (UT) & T$_{SL}$ & Object & PID & Date (UT) & T$_{SL}$ \\
\\
\hline
ARP 220 & 105 & 2004 Feb 29 & 3$\times$14 & IRAS 14060+2919 & 2306 & 2004 Jul 16 & 2$\times$60 \\
IRAS 00091$-$0738 & 3187 & 2005 Jun 30 & 2$\times$60 & IRAS 14197+0813 & 3187 & 2005 Feb 13 & 2$\times$60 \\
IRAS 00188$-$0856 & 105 & 2003 Dec 17 & 2$\times$60 & IRAS 14252$-$1550 & 2306 & 2004 Jul 17 & 2$\times$60 \\
IRAS 00456$-$2904 & 3187 & 2005 Jul 14 & 2$\times$60 & IRAS 14348$-$1447 & 105 & 2004 Feb 07 & 1$\times$60 \\
IRAS 00482$-$2721 & 3187 & 2005 Jul 07 & 2$\times$60 & IRAS 15130$-$1958 & 3187 & 2005 Mar 15 & 2$\times$60 \\
IRAS 01003$-$2238 & 105 & 2004 Jan 04 & 1$\times$60 & IRAS 15206+3342 & 105 & 2004 Jun 24 & 1$\times$60 \\
IRAS 01166$-$0844$^a$ & 3187 & 2005 Jan 03 & 2$\times$60 & IRAS 15225+2350 & 2306 & 2005 Feb 07 & 2$\times$60 \\
IRAS 01298$-$0744 & 105 & 2005 Jul 14 & 2$\times$60 & IRAS 15250+3609 & 105 & 2004 Mar 04 & 3$\times$14 \\
IRAS 01569$-$2939 & 2306 & 2004 Jul 18 & 2$\times$60 & IRAS 15462$-$0450 & 105 & 2004 Mar 02 & 1$\times$60 \\
IRAS 02021$-$2103 & 3187 & 2005 Jan 15 & 2$\times$60 & IRAS 16090$-$0139 & 105 & 2005 Aug 05 & 1$\times$60 \\
IRAS 02411+0353 & 2306 & 2005 Jan 14 & 2$\times$60 & IRAS 16156+0146 & 3187 & 2005 Mar 15 & 2$\times$60 \\
IRAS 03250+1606 & 3187 & 2005 Feb 11 & 2$\times$60 & IRAS 16468+5200$^c$ & 2306 & 2004 Jul 14 & 4$\times$60 \\
IRAS 04103$-$2838 & 3187 & 2005 Feb 10 & 2$\times$60 & IRAS 16474+3430 & 2306 & 2004 Jul 14 & 2$\times$60 \\
IRAS 05189$-$2524 & 105 & 2004 Mar 22 & 3$\times$14 & IRAS 16487+5447 & 2306 & 2004 Jul 17 & 2$\times$60 \\
IRAS 07598+6508 & 105 & 2004 Feb 29 & 3$\times$14 & IRAS 17028+5817 & 2306 & 2004 Jul 17 & 2$\times$60 \\
IRAS 08559+1053$^b$ & 666 & 2003 Nov 23 & 2$\times$60 & IRAS 17044+6720 & 2306 & 2004 Jul 17 & 2$\times$60 \\
IRAS 08572+3915 & 105 & 2004 Apr 15 & 3$\times$14 & IRAS 17179+5444 & 105 & 2004 Apr 17 & 2$\times$60 \\
IRAS 09039+0503 & 3187 & 2005 Apr 18 & 2$\times$60 & IRAS 17208$-$0014 & 105 & 2004 Mar 27 & 3$\times$14 \\
IRAS 09116+0334 & 2306 & 2005 Apr 21 & 2$\times$60 & IRAS 19254$-$7245 & 105 & 2005 May 30 & 3$\times$14 \\
IRAS 09539+0857 & 3187 & 2005 Jun 05 & 2$\times$60 & IRAS 20100$-$4156 & 105 & 2004 Apr 13 & 1$\times$60 \\
IRAS 10190+1322$^c$ & 3187 & 2005 May 22 & 4$\times$60 & IRAS 20414$-$1651 & 105 & 2004 May 14 & 1$\times$60 \\
IRAS 10378+1109 & 105 & 2005 Jun 08 & 2$\times$60 & IRAS 20551$-$4250 & 105 & 2004 May 14 & 2$\times$14 \\
IRAS 10485$-$1447 & 3187 & 2005 May 23 & 2$\times$60 & IRAS 21208$-$0519$^d$ & 3187 & 2004 Nov 13 & 2$\times$60 \\
IRAS 10494+4424 & 2306 & 2004 Nov 17 & 2$\times$60 & IRAS 21219$-$1757 & 3187 & 2004 Nov 16 & 2$\times$60 \\
IRAS 11095$-$0238 & 105 & 2005 Jun 07 & 2$\times$60 & IRAS 21329$-$2346 & 3187 & 2004 Nov 16 & 2$\times$60 \\
IRAS 11130$-$2659 & 2306 & 2005 Jul 12 & 2$\times$60 & IRAS 22206$-$2715 & 3187 & 2004 Nov 15 & 2$\times$60 \\
IRAS 11387+4116 & 2306 & 2005 Jan 11 & 2$\times$60 & IRAS 22491$-$1808 & 105 & 2004 Jun 24 & 1$\times$60 \\
IRAS 11506+1331 & 3187 & 2005 May 25 & 2$\times$60 & IRAS 23128$-$5919 & 105 & 2004 May 11 & 3$\times$14 \\
IRAS 12072$-$0444 & 105 & 2004 Jan 06 & 1$\times$60 & IRAS 23234+0946 & 3187 & 2004 Dec 13 & 2$\times$60 \\
IRAS 12112+0305 & 105 & 2004 Jan 04 & 3$\times$14 & IRAS 23327+2913 & 2306 & 2004 Dec 08 & 2$\times$60 \\
IRAS 12127$-$1412 & 3187 & 2005 Jun 30 & 2$\times$60 & MRK 231 & 105 & 2004 Apr 14 & 2$\times$14 \\
IRAS 12359$-$0725 & 2306 & 2005 Jun 30 & 2$\times$60 & MRK 273 & 105 & 2004 Apr 14 & 2$\times$14 \\
IRAS 13335$-$2612 & 3187 & 2005 Feb 15 & 2$\times$60 & NGC 6240 & 105 & 2004 Mar 04 & 2$\times$14 \\
IRAS 13454$-$2956$^c$ & 3187 & 2005 Jul 14 & 4$\times$60 & 4C +12.50 & 105 & 2004 Jan 07 & 3$\times$14 \\
IRAS 13509+0442 & 2306 & 2004 Jul 17 & 2$\times$60 & UGC 5101 & 105 & 2004 Mar 23 & 3$\times$14 \\
IRAS 13539+2920 & 2306 & 2005 Feb 07 & 2$\times$60 & \\
\hline
\end{tabular}
\end{scriptsize}
\end{center}
\begin{flushleft}
\textit{Notes.} $^a$~Southern nucleus. The northern one is also observed but is much fainter. 
$^b$~Observation performed during the in-orbit checkout phase 
(IRS campaign P: PID 666, PI J.R.Houck). $^c$~Both nuclei are observed but apparently not 
completely resolved. $^d$~Northern nucleus. The southern one is also observed but is much fainter.
\end{flushleft}
\end{table*}

The spectroscopic observations of all the sources in our sample were 
obtained with the \textit{Spitzer}-IRS low-resolution modules, within 
three different programs devoted to buried activity in ULIRGs and massive 
mergers\footnote{The only exception is IRAS~08559+1053, as mentioned.}: 
the program IDs are 105, 2306 and 3187, and the principal investigators 
are J.R.Houck, M.Imanishi and S.Veilleux, respectively. The details concerning 
each observation are listed in Tab.\ref{t2}. The local ULIRGs of our sample are 
\textit{bright} objects, and their signal can be easily distinguished from 
the background emission, mostly due to zodiacal light. After having verified 
that a refined reduction of the data was not necessary for our purposes, we 
decided to perform our analysis beginning from the coadded images provided by 
the \textit{Spitzer Science Center}. A coadded image is obtained as the average 
over multiple telescope pointings. At this stage the individual snapshots have 
been already processed with the default pipeline (versions 13.0 and upgrades), 
that among other basic operations includes the linearization and the fitting of 
the signal ramp, the dark subtraction and the flat-fielding. Since each observation 
in staring mode consists of two exposures with different position of the source along 
the slit, we have subtracted the background emission by taking the difference between 
the couple of images in the nodding cycle. The spectra have been extracted following 
the standard steps for point-like sources within the software \texttt{SPICE}. 
The flux uncertainties have been computed considering the typical poissonian 
distribution for source and background counts. The latter are estimated from 
the companion coadded images, since the off-source subslit provides 
the background of the nodded on-source observation. \\ 
While the wavelength calibration is not a crucial issue in this work, a precise 
flux calibration is required to derive a reliable estimate of the relative AGN/SB 
contribution to the bolometric luminosity (see Section~\ref{ae}). In the older 
versions of the processing pipeline the accuracy of the absolute flux calibration for 
the \textit{Spitzer}-IRS low-resolution orders was quoted to be about 20 per cent. 
Even with such a coarse evaluation our final results are not substantially modified, 
hence the \textit{Spitzer} calibration as been adopted. We have anyway checked its 
reliability against the photometric data provided by IRAS. For 53 sources 
(corresponding to 75 per cent of our sample) the 12~$\mu$m IRAS flux density is only 
an upper limit, which is always higher than the \textit{Spitzer}-IRS equivalent. 
For 16 sources not even the IRAS flux at 25~$\mu$m is available, while in all 
the remaining cases the \textit{Spitzer} measure turns out to be fully consistent 
with the IRAS one (it is indeed slightly lower on average, by $\sim$20 per cent at 
most). Even if the analysis is restricted to a narrow spectral range and only the 
Short-Low (SL) orders are involved, we performed the extraction of the Long-Low (LL) 
orders as well, so that we have the entire $\sim$5--35~$\mu$m spectra of more than 
60 sources. Only in a few cases a slight rescaling is necessary for a smooth connection 
between SL1 and LL2 orders, and this can not introduce any systematic effect on the 
measured 25~$\mu$m flux. Aperture losses at long wavelengths can indeed occur when 
comparing the narrow-slit IRS spectroscopy to the large-aperture IRAS photometry. 
Whatever the right explanation may be, we conclude that the \textit{Spitzer}-IRS 
calibration at 5--8~$\mu$m is fully reliable, and that the normalization of our 
spectra to match the IRAS fluxes, however not possible for the whole sample, is 
not even required.

\section{AGN/SB bolometric contribution}
\label{ae}

\begin{figure}
\includegraphics[width=8.5cm]{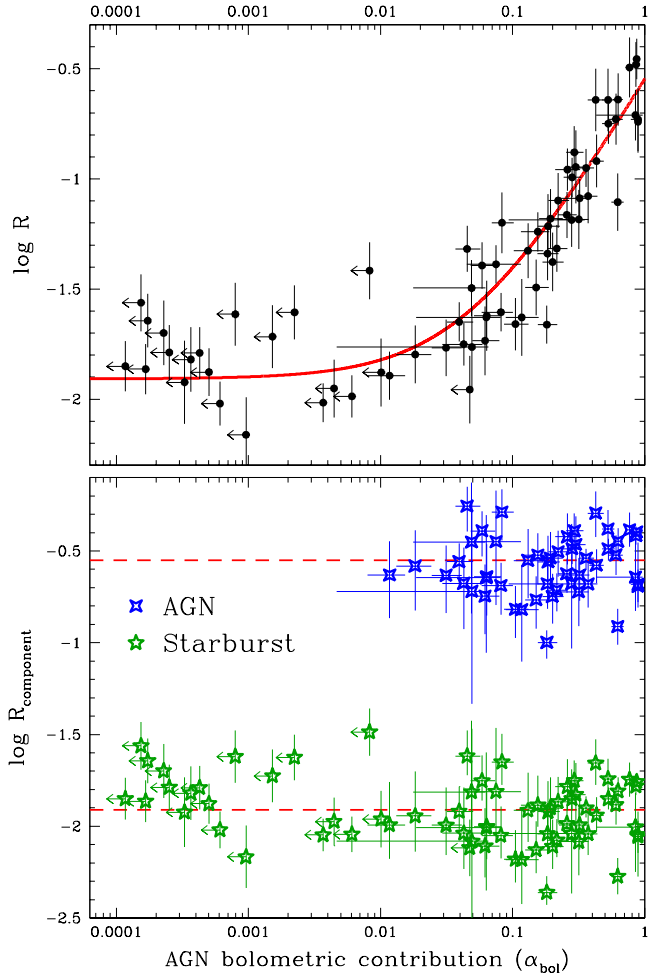}
\caption{\textit{(a)} Ratio $R$ between intrinsic 6~$\mu$m and bolometric luminosity 
versus the AGN bolometric contribution $\alpha_\mathit{bol}$. The error bars of $R$ are 
mainly due to the uncertainties in the total IR flux $F_{IR}$. The \textit{red 
solid curve} traces the best fit of the $R$--$\alpha_6$ relation from Eq.\ref{ra}. It is 
worth noting that the sources follow a very regular pattern and no evidence of outliers 
is found. \textit{(b)} Same as above, with the ratios for the AGN and SB components 
plotted separately. The \textit{red dashed lines} mark the best values of $R^\mathit{agn}$ 
and $R^\mathit{sb}$. The agreement among the SB components at different values of 
$\alpha_\mathit{bol}$ proves the reliability of our method (see the discussion in 
the text).}
\label{np}
\end{figure}

The use of spectral templates to model the intrinsic AGN and SB components 
significantly reduces the number of degrees of freedom in our analytical description. 
The decomposition method has already been introduced and discussed in detail in 
Paper~I, nevertheless it is useful to examine again the main steps. 
By defining the 6~$\mu$m--normalized AGN and SB templates $u_\nu^\mathit{agn}$ 
and $u_\nu^\mathit{sb}$, the observed 5--8~$\mu$m ULIRG emission can be 
parameterized as follows: 
\begin{equation}
f_\nu^\mathit{obs}(\lambda)=f_6^\mathit{int}\left[(1-\alpha_6)u_\nu^\mathit{sb}+
\alpha_6 u_\nu^\mathit{agn} e^{-\tau(\lambda)}\right], 
\end{equation}
where $\alpha_6$ is the AGN contribution to the intrinsic (de-absorbed) flux 
density $f_6^\mathit{int}$. Along with the optical depth to the active nucleus 
$\tau_6=\tau$(6~$\mu$m) and apart from the flux normalization, $\alpha_6$ is 
the only degree of freedom in our model. The best fits have been found via 
the minimization of the $\chi^2$. In spite of the large differences observed 
in our ULIRG spectra, this simple model always provides a good match to the data. 
Though the best fits have usually a reduced $\chi^2\ga$2 and are not formally 
acceptable in a statistical sense (due to the use of templates that ignores minor 
features, and the small \textit{Spitzer}-IRS error bars) both the aromatic emission  
and the continuum shape are well reproduced, with residuals smaller than 10 per cent 
at all wavelengths. \\
The fitting procedure allows the determination of $f_6^\mathit{int}$ as well, 
hence we can compute the ratio between the absorption-corrected 6~$\mu$m 
luminosity and the bolometric luminosity of each source, defined as 
\begin{equation}
R=\left(\frac{\nu_6 f_6^\mathit{int}}{F_\mathit{IR}}\right),
\label{eq}
\end{equation}
where $F_\mathit{IR}$ is the total IR flux introduced in Eq.(\ref{e1}). 
Since the equivalent quantities for pure AGN and pure starbursts 
(hereafter indicated respectively as $R^\mathit{agn}$ and $R^\mathit{sb}$) 
are widely different from each other, $R$ is itself an indicator of the 
significance of AGN activity within composite sources, and can be 
used both to test the consistency of our decomposition method and to 
assess the relative AGN/SB contribution in terms of bolometric luminosity. 
By rendering explicit the AGN and SB components within Eq.(\ref{eq}) respectively 
as $\alpha_6 f_6^\mathit{int}$ and $(1-\alpha_6)f_6^\mathit{int}$, and 
decomposing $F_\mathit{IR}$ as $F_\mathit{IR}^\mathit{agn}+F_\mathit{IR}^\mathit{sb}$, 
the dependence of $R$ on $\alpha_6$, $R^\mathit{agn}$ and $R^\mathit{sb}$ 
can be brought out through a simple algebra: 
\begin{equation}
R=\frac{R^\mathit{agn}R^\mathit{sb}}{\alpha_6 R^\mathit{sb}+(1-\alpha_6) R^\mathit{agn}}.
\label{ra}
\end{equation}
This theoretical $R$--$\alpha_6$ relation has been fitted to our data, 
with $R^\mathit{agn}$ and $R^\mathit{sb}$ as floating variables. 
In Paper~I we obtained that $R^\mathit{agn}/R^\mathit{sb}\sim$28; 
we have anyway decided to repeat the evaluation of these critical 
parameters, because some fits have been improved with respect to the 
previous work and three extra sources have been added to the sample. 
The new results are in good agreement with the previous ones: 
\begin{displaymath}
\log R^\mathit{agn}=-0.55^{+0.06}_{-0.07} \hspace*{10pt} \textrm{and} \hspace*{10pt} \log R^\mathit{sb}=-1.91^{+0.02}_{-0.02}.
\end{displaymath}
Taking $R^\mathit{agn}/R^\mathit{sb}\sim$23, we are able to provide a 
quantitative estimate of the AGN contribution 
(${\alpha_\mathit{bol}=F_\mathit{IR}^\mathit{agn}/F_\mathit{IR}}$) to 
the bolometric luminosity of each source, as 
\begin{equation}
\alpha_\mathit{bol}=\frac{\alpha_6}{\alpha_6+(R^\mathit{agn}/R^\mathit{sb})(1-\alpha_6)}.
\label{ab}
\end{equation}
The values of $\alpha_\mathit{bol}$ are listed in Tab.\ref{t3}. Incidentally, 
it should be noted that a value $\alpha_\mathit{bol} \approx 1$ is not reached even 
in optically bright quasars, since the contamination from star formation activity is 
always present and non-negligible. In Fig.\ref{np}a we show the relation between $R$ 
and $\alpha_\mathit{bol}$, which can be expressed in the neat and manifest form 
$R=\alpha_\mathit{bol}R^\mathit{agn}+(1-\alpha_\mathit{bol})R^\mathit{sb}$  
by inverting Eq.\ref{ab}. \\ 
The regularity of the pattern shaped by the location of the sources in this plot 
is in itself a qualitative validation of our method. We have anyway performed a 
further test of self-consistency, by computing for each source the following quantities: 
\begin{equation}
\widehat R^\mathit{agn}=\left(\frac{\nu_6 \alpha_6 f_6^\mathit{int}}{\alpha_\mathit{bol} F_\mathit{IR}}\right)  \ \ \ \textrm{and} \ \ \
\widehat R^\mathit{sb}=\left[\frac{\nu_6 (1-\alpha_6) f_6^\mathit{int}}{(1-\alpha_\mathit{bol}) F_\mathit{IR}}\right].
\end{equation}
Although this argument may appear as fully circular, it indeed provides a strong 
confirmation of our approach. In fact, as shown in Fig.\ref{np}b, the ratios 
$\widehat R^\mathit{sb}$ for the SB component in composite sources (which depend 
on our AGN/SB decomposition) agree within the dispersion with those for SB-dominated 
ULIRGs (which conversely are directly measured). Such agreement would be lost both 
in case of missed AGN detections and in case of fake AGN detections, since the 
global IR flux is sensitive to the actual AGN contribution. \\ 
The present method is then reliable in detecting genuine AGN components and in 
extrapolating the average AGN/SB contribution to the bolometric luminosities. 
Moreover, in spite of its simplicity, it proves to be effective when compared 
to other diagnostics at different wavelengths, as discussed in the next Section.
It is important however to underline that the scatter in $\widehat R^\mathit{agn}$ 
and $\widehat R^\mathit{sb}$ is significantly larger than the uncertainty on the best 
values of $R^\mathit{agn}$ and $R^\mathit{sb}$. The reduced $\chi^2$ of the $R$--$\alpha_6$ 
relation varies slowly within a region of the parameter space that 
roughly corresponds to the dispersion around the best fit in Fig.\ref{np}a. We then emphasize 
that such a dispersion ($\sim$0.3 dex, nearly constant with respect to $\alpha_\mathit{bol}$) 
is to be considered the real uncertainty in the 6~$\mu$m to bolometric ratios for the 
AGN and SB components of individual objects (as opposed to the ensemble averages). 
This can slightly affect the estimates of $\alpha_\mathit{bol}$, increasing the statistical 
errors reported in Tab.\ref{t3} as well. Nevertheless the results are precise 
enough to establish the magnitude of the AGN/SB contribution both to the luminosity of 
individual sources and to the overall energy output of local ULIRGs. \\

\begin{table*}
\begin{center}
\caption{Spectral parameters and AGN contribution for the 71 ULIRGs in our sample. 
All the optical depths refer to the AGN component: a \textit{colon} indicates the 
upper limits evaluated from the residuals to the best fit. 
$\alpha_6$: AGN contribution to the intrinsic (i.e. de-absorbed) continuum emission 
at 6~$\mu$m (in per cent). $\tau_6$: Optical depth to the AGN continuum at 6~$\mu$m 
according to the power-law prescription for reddening. $\alpha_\mathit{bol}$: 
AGN contribution to the bolometric luminosity (in per cent). These entries are only 
affected by the statistical uncertainty both in the flux amplitude of the AGN/SB
components and in the average ratios $R^\mathit{agn}$ and $R^\mathit{sb}$. 
The systematic effects are discussed and quantified in the text. $\tau_\mathit{ice}$: 
Optical depth of the $\sim$6.0~$\mu$m absorption feature due to water ice. 
$\tau_\mathit{hac,1}$, $\tau_\mathit{hac,2}$: Optical depths of the companion features 
attributed to hydrogenated amorphous carbons, centred at $\sim$6.85 and $\sim$7.25~$\mu$m, 
respectively. $\alpha_6$*, $\tau_6$*, $\alpha_\mathit{bol}$*: Same as $\alpha_6$, $\tau_6$ 
and $\alpha_\mathit{bol}$, obtained by assuming a spectral slope of 0.7 (instead of 1.5) 
for the AGN template (see Section~\ref{dis}). $V/L/X$: Spectral classification in the visible, 
\textit{L}-band and hard X-rays. $\star$: Starburst; $\otimes$: LINER; $\bullet$: AGN detection; 
$\odot$: AGN detection, tentative.}
\label{t3}
\begin{scriptsize}
\begin{tabular}{lcccccccccc}
\hline \hline
Source & $\alpha_6$ & $\tau_6$ & $\alpha_\mathit{bol}$ & $\tau_\mathit{ice}$ & $\tau_\mathit{hac,1}$ & 
$\tau_\mathit{hac,2}$ & $\alpha_6$* & $\tau_6$* & $\alpha_\mathit{bol}$* & $V/L/X$ \\
\hline
ARP 220 & 84$\pm$1 & 1.47$\pm$0.01 & 18$^{+4}_{-3}$ & $<2.13$ & 0.29$\pm$0.01 & $<0.21$: & 90$\pm$1 & 
2.00$\pm$0.01 & 20$^{+4}_{-3}$ & \small $\otimes^1 \star^4 \odot^8$ \\
IRAS 00091$-$0738 & 97$\pm$1 & 2.30$\pm$0.04 & 60$^{+8}_{-7}$ & 0.88$\pm$0.10 & 0.57$\pm$0.08 & 0.34$\pm$0.04 & 
98$\pm$1 & 2.80$\pm$0.03 & 62$\pm$6 & \small $\star^1 - -$ \\
IRAS 00188$-$0856$^a$ & 93$\pm$1 & 0.37$\pm$0.04 & 37$^{+6}_{-5}$ & $<1.18$ & $<0.67$ & $<0.66$ & 93$\pm$1 & 
0.42$\pm$0.01 & 29$^{+5}_{-4}$ & \small $\otimes^1 \odot^4 -$ \\
IRAS 00456$-$2904 & $<1.0$ & $-$ & $<0.05$ & $-$ & $-$ & $-$ & $<1.0$ & $-$ & $<0.03$ & 
\small $\star^1 - -$ \\
IRAS 00482$-$2721$^b$ & $<54$ & $<0.04$ & $<4.8$ & $<0.01$ & $<0.96$: & $<0.79$: & $<38$ & $<0.07$ & $<1.7$ & 
\small $\otimes^1 - -$ \\
IRAS 01003$-$2238 & 96$\pm$1 & 1.58$\pm$0.02 & 53$^{+6}_{-5}$ & 0.18$\pm$0.04 & 0.14$\pm$0.01 & $<0.01$ & 
98$\pm$1 & 2.12$\pm$0.02 & 56$^{+6}_{-5}$ & \small $\odot^1 - -$ \\
IRAS 01166$-$0844 & $>99$ & 2.35$^{+0.23}_{-0.08}$ & 89$^{+6}_{-11}$ & 0.41$^{+0.36}_{-0.33}$ & 
0.70$^{+0.38}_{-0.35}$ & 0.40$^{+0.19}_{-0.18}$ & $>99$ & 2.92$\pm$0.04 & 90$\pm$3 & 
\small $\star^1 - -$ \\
IRAS 01298$-$0744 & $>98$ & 1.79$\pm$0.02 & 77$^{+5}_{-6}$ & 0.91$\pm$0.05 & 0.47$\pm$0.03 & 
0.29$\pm$0.02 & $>99$ & 2.29$\pm$0.02 & 78$\pm$5 & \small $\star^1 - -$ \\
IRAS 01569$-$2939 & 85$\pm$1 & 1.13$^{+0.02}_{-0.04}$ & 19$^{+5}_{-3}$ & 0.63$^{+0.09}_{-0.07}$ & 0.31$\pm$0.09 & 
0.31$\pm$0.05 & 91$\pm$1 & 1.66$\pm$0.05 & 22$^{+5}_{-4}$ & \small $\star^1 - -$ \\
IRAS 02021$-$2103 & 78$\pm$2 & 1.91$\pm$0.10 & 13$^{+4}_{-3}$ & $<0.01$ & $<0.40$: & $<0.18$: & 86$\pm$1 & 
2.47$\pm$0.11 & 15$^{+4}_{-3}$ & \small $- - -$ \\
IRAS 02411+0353$^c$ & $<17$ & $-$ & $<0.9$ & $-$ & $-$ & $-$ & 30$\pm$2 & 0.77$\pm$0.07 & 1.2$^{+0.4}_{-0.3}$ & 
\small $\star^1 - -$ \\
IRAS 03250+1606 & $<3.4$ & $<0.11$ & $<0.2$ & $-$ & $-$ & $-$ & $<3.9$ & $<0.14$ & $<0.2$ & 
\small $\otimes^1 \odot^4 -$ \\
IRAS 04103$-$2838 & 59$\pm$1 & 0.14$\pm$0.02 & 5.9$^{+1.5}_{-1.1}$ & $<0.01$ & $<0.01$ & $<0.01$ & 70$\pm$1 & 
0.63$\pm$0.02 & 6.3$^{+1.6}_{-1.1}$ & \small $\otimes^1 - \bullet^9$ \\
IRAS 05189$-$2524 & 92$^{+1}_{-3}$ & $<0.01$ & 32$^{+6}_{-10}$ & $<0.15$ & 0.04$\pm$0.01 & $<0.01$ & 94$\pm$1 & 
0.43$\pm$0.01 & 30$^{+5}_{-4}$ & \small $\bullet^1 \bullet^4 \bullet^{10}$ \\
IRAS 07598+6508$^d$ & $>99$ & $<0.06$ & 86$\pm$4 & $<0.01$ & $<0.01$ & $<0.02$ & $>97$ & $<0.01$ & 
94$^{+4}_{-44}$ & \small $\bullet^2 \bullet^4 \odot^{11}$ \\
IRAS 08559+1053 & 68$\pm$1 & $<0.01$ & 8.3$^{+1.9}_{-1.4}$ & $<0.01$ & $<0.16$ & $<0.12$: & 65$^{+1}_{-12}$ & 
$<0.01$ & 5.0$^{+1.2}_{-2.4}$ & \small $\bullet^1 \bullet^4 -$ \\
IRAS 08572+3915 & $>99$ & 0.44$\pm$0.01 & 87$\pm$3 & 0.16$\pm$0.02 & 0.24$\pm$0.01 & 0.10$\pm$0.01 & $>99$ & 
0.91$\pm$0.01 & 81$\pm$4 & \small $\otimes^1 \bullet^4 -$ \\
IRAS 09039+0503 & 61$\pm$1 & 0.71$\pm$0.04 & 6.4$^{+1.8}_{-1.3}$ & $<1.33$ & $<0.01$ & $<0.38$: & 72$\pm$1 & 
1.20$\pm$0.04 & 6.9$^{+1.9}_{-1.3}$ & \small $\otimes^1 \odot^4 -$ \\
IRAS 09116+0334 & $<1.8$ & $-$ & $<0.08$ & $-$ & $-$ & $-$ & $<0.9$ & $-$ & $<0.03$ & 
\small $\otimes^1 \odot^4 -$ \\
IRAS 09539+0857 & 91$\pm$1 & 1.85$\pm$0.03 & 30$^{+6}_{-4}$ & $<1.27$ & 0.31$\pm$0.03 & 0.28$\pm$0.02 & 
94$\pm$1 & 2.38$\pm$0.03 & 33$^{+6}_{-5}$ & \small $\otimes^1 \star^4 -$ \\
IRAS 10190+1322 & $<0.3$ & $-$ & $<0.02$ & $-$ & $-$ & $-$ & $<0.3$ & $-$ & $<0.01$ & 
\small $\star^1 \star^4 -$ \\
IRAS 10378+1109 & 80$\pm$1 & 0.21$\pm$0.02 & 15$^{+4}_{-3}$ & $<2.56$ & 0.43$\pm$0.03 & 0.48$\pm$0.03 & 
87$\pm$1 & 0.70$\pm$0.02 & 16$^{+4}_{-3}$ & \small $\otimes^1 \odot^4 -$ \\
IRAS 10485$-$1447 & 60$\pm$1 & 0.16$\pm$0.03 & 6.2$^{+1.7}_{-1.2}$ & $<2.00$ & 0.50$\pm$0.07 & $<0.34$: & 
70$\pm$1 & 0.61$\pm$0.04 & 6.2$^{+1.7}_{-1.2}$ & \small $\otimes^1 \odot^4 -$ \\
IRAS 10494+4424 & $<0.4$ & $-$ & $<0.02$ & $-$ & $-$ & $-$ & $<0.4$ & $-$ & $<0.01$ & 
\small $\otimes^1 \odot^4 -$ \\
IRAS 11095$-$0238 & 97$\pm$1 & 1.56$\pm$0.01 & 63$\pm$5 & 0.45$\pm$0.02 & 0.59$\pm$0.02 & 0.38$\pm$0.01 & 
98$\pm$1 & 2.05$^{+0.08}_{-0.04}$ & 62$^{+11}_{-8}$ & \small $\otimes^1 \star^4 -$ \\
IRAS 11130$-$2659 & 84$\pm$1 & 1.15$\pm$0.02 & 19$^{+4}_{-3}$ & 0.37$\pm$0.05 & 0.70$\pm$0.05 & 0.40$\pm$0.07 & 
90$\pm$1 & 1.67$\pm$0.02 & 21$\pm$4 & \small $\otimes^1 - -$ \\
IRAS 11387+4116 & $<0.8$ & $-$ & $<0.04$ & $-$ & $-$ & $-$ & $<0.7$ & $-$ & $<0.02$ & 
\small $\star^1 \star^4 -$ \\
IRAS 11506+1331 & 54$^{+1}_{-21}$ & $<0.01$ & 4.9$^{+1.3}_{-3.1}$ & $<0.29$ & 0.32$^{+0.02}_{-0.20}$ & 
$<0.01$ & 57$\pm$1 & 0.18$\pm$0.02 & 3.6$^{+0.9}_{-0.7}$ & \small $\star^1 \odot^4 -$ \\
IRAS 12072$-$0444 & 95$\pm$1 & 1.07$\pm$0.01 & 43$^{+6}_{-5}$ & 0.28$\pm$0.02 & 0.26$\pm$0.01 & 
0.19$\pm$0.01 & 97$\pm$1 & 1.58$\pm$0.01 & 45$^{+6}_{-5}$ & \small $\bullet^1 \bullet^4 -$ \\
IRAS 12112+0305 & $<13$ & $<0.07$ & $<0.7$ & $-$ & $-$ & $-$ & $<9.7$ & $<0.04$ & $<0.4$ & 
\small $\otimes^1 \star^5 \star^{12}$ \\
IRAS 12127$-$1412$^e$ & $>99$ & $<0.04$ & 89$^{+3}_{-4}$ & $<0.44$ & $<0.40$ & $<0.38$ & $>99$ & 0.37$\pm$0.01 & 
84$^{+3}_{-4}$ & \small $\otimes^1 \bullet^4 -$ \\
IRAS 12359$-$0725 & 54$^{+2}_{-43}$ & $<0.01$ & 4.9$^{+1.6}_{-4.4}$ & $<0.37$ & $<0.48$ & $<0.38$ & 
53$^{+1}_{-8}$ & $<0.01$ & 3.1$^{+0.8}_{-1.2}$ & \small $\otimes^1 \odot^4 -$ \\
IRAS 13335$-$2612 & $<0.6$ & $-$ & $<0.03$ & $-$ & $-$ & $-$ & $<0.5$ & $-$ & $<0.02$ & 
\small $\otimes^1 - -$ \\
IRAS 13454$-$2956 & 61$^{+1}_{-27}$ & $<0.01$ & 6.3$^{+1.8}_{-4.5}$ & $<0.01$ & $<0.16$: & $<0.09$: & 
59$\pm$1 & 0.07$\pm$0.04 & 4.0$^{+1.1}_{-0.8}$ & \small $\bullet^1 - -$ \\
IRAS 13509+0442 & $<0.6$ & $-$ & $<0.03$ & $-$ & $-$ & $-$ & $<0.6$ & $-$ & $<0.02$ & 
\small $\star^1 \star^4 -$ \\
IRAS 13539+2920 & $<0.4$ & $-$ & $<0.02$ & $-$ & $-$ & $-$ & $<0.4$ & $-$ & $<0.02$ & 
\small $\star^1 \star^4 -$ \\
IRAS 14060+2919 & $<0.4$ & $-$ & $<0.02$ & $-$ & $-$ & $-$ & $<0.4$ & $-$ & $<0.01$ & 
\small $\star^1 \star^4 -$ \\
IRAS 14197+0813$^f$ & 75$\pm$2 & 2.10$\pm$0.14 & 12$^{+4}_{-3}$ & $<1.61$ & $<0.01$ & $<0.01$ & 84$\pm$2 & 
2.63$\pm$0.15 & 13$^{+4}_{-3}$ & \small $- - -$ \\
IRAS 14252$-$1550 & $<20$ & $<0.04$ & $<1.1$ & $-$ & $-$ & $-$ & $<24$ & $<0.31$ & $<0.9$ & 
\small $\otimes^1 \star^4 -$ \\
IRAS 14348$-$1447 & 51$^{+3}_{-5}$ & $<0.09$ & 4.3$^{+1.7}_{-1.3}$ & $<1.94$ & 0.57$^{+0.32}_{-0.36}$ & 
0.62$^{+0.18}_{-0.26}$ & 59$\pm$3 & 0.25$\pm$0.10 & 3.9$^{+1.5}_{-1.0}$ & 
\small $\otimes^1 \star^5 \star^{12}$ \\
IRAS 15130$-$1958 & 91$^{+1}_{-3}$ & $<0.01$ & 32$^{+6}_{-10}$ & $<0.08$ & $<0.01$ & $<0.01$ & 94$\pm$1 & 
0.42$\pm$0.01 & 30$^{+5}_{-4}$ & \small $\bullet^1 \bullet^4 -$ \\
IRAS 15206+3342 & 52$\pm$1 & 0.30$\pm$0.02 & 4.5$^{+1.2}_{-0.8}$ & $<0.01$ & $<0.01$ & $<0.01$ & 65$\pm$1 & 
0.83$\pm$0.02 & 5.1$^{+1.3}_{-0.9}$ & \small $\star^1 \star^4 -$ \\
IRAS 15225+2350 & 87$\pm$1 & 0.74$\pm$0.02 & 22$^{+5}_{-3}$ & 0.92$\pm$0.05 & 0.35$\pm$0.02 & 
0.22$\pm$0.01 & 92$\pm$1 & 1.26$\pm$0.02 & 24$^{+5}_{-4}$ & \small $\star^1 \odot^4 -$ \\
IRAS 15250+3609 & 96$\pm$1 & 1.06$\pm$0.01 & 53$\pm$6 & 1.38$\pm$0.03 & 0.60$\pm$0.03 & 
0.41$\pm$0.02 & 98$\pm$1 & 1.58$\pm$0.01 & 55$^{+6}_{-5}$ & \small $\otimes^2 - \star^{12}$ \\
IRAS 15462$-$0450 & 90$^{+1}_{-16}$ & $<0.01$ & 28$^{+7}_{-19}$ & $<0.01$ & $<0.07$: & 
0.09$^{+0.01}_{-0.05}$ & 89$\pm$1 & 0.22$\pm$0.01 & 19$^{+4}_{-3}$ & 
\small $\bullet^1 \bullet^4 -$ \\
IRAS 16090$-$0139 & 89$\pm$1 & 0.69$\pm$0.01 & 26$\pm$4 & 0.90$\pm$0.03 & 0.57$\pm$0.02 & 
0.36$\pm$0.01 & 93$\pm$1 & 1.21$\pm$0.01 & 28$^{+5}_{-4}$ & \small $\otimes^1 \odot^4 -$ \\
IRAS 16156+0146 & 94$\pm$1 & 0.61$\pm$0.02 & 43$^{+6}_{-5}$ & 0.48$\pm$0.04 & 0.24$\pm$0.01 & 
0.11$\pm$0.01 & 96$\pm$1 & 1.09$\pm$0.01 & 43$^{+6}_{-5}$ &\small $\bullet^1 - -$ \\
IRAS 16468+5200 & 85$\pm$1 & 0.77$\pm$0.02 & 20$^{+4}_{-3}$ & $<1.50$ & 0.94$\pm$0.08 & 
0.60$\pm$0.04 & 91$\pm$1 & 1.30$\pm$0.02 & 22$^{+5}_{-3}$ & \small $\otimes^1 \star^4 -$ \\
IRAS 16474+3430 & $<4.9$ & $-$ & $<0.3$ & $-$ & $-$ & $-$ & $<5.5$ & $<0.01$ & $<0.2$ & 
\small $\star^1 \odot^4 -$ \\
IRAS 16487+5447$^g$ & 21$^{+1}_{-3}$ & $<0.04$ & 1.2$\pm$0.4 & $-$ & $-$ & $-$ & 31$\pm$2 & 
0.47$\pm$0.08 & 1.2$^{+0.5}_{-0.3}$ & \small $\otimes^1 \odot^4 -$ \\
IRAS 17028+5817 & $<1.2$ & $-$ & $<0.06$ & $-$ & $-$ & $-$ & $<1.1$ & $-$ & $<0.04$ & 
\small $\otimes^1 \odot^4 -$ \\
IRAS 17044+6720 & 91$\pm$1 & 0.32$\pm$0.01 & 29$^{+5}_{-4}$ & 0.09$\pm$0.02 & $<0.01$ & 
$<0.01$ & 94$\pm$1 & 0.82$\pm$0.01 & 31$^{+5}_{-4}$ & \small $\otimes^1 \bullet^4 -$ \\
IRAS 17179+5444 & 84$\pm$1 & 0.31$\pm$0.02 & 18$^{+4}_{-3}$ & $<0.01$ & $<0.01$ & $<0.01$ & 
89$\pm$1 & 0.81$\pm$0.02 & 19$^{+4}_{-3}$ & \small $\bullet^1 \bullet^4 -$ \\
IRAS 17208$-$0014 & $<7.9$ & $-$ & $<0.4$ & $-$ & $-$ & $-$ & $<7.5$ & $-$ & $<0.3$ & 
\small $\otimes^3 \star^5 \star^{12}$ \\
IRAS 19254$-$7245$^h$ & 89$\pm$1 & 0.21$\pm$0.08 & 26$^{+6}_{-5}$ & 0.21$^{+0.17}_{-0.16}$ & 
0.21$\pm$0.04 & 0.10$\pm$0.02 & 88$\pm$1 & 0.09$\pm$0.01 & 17$\pm$3 & 
\small $\bullet^3 \bullet^5 \bullet^{12}$ \\
\hline
\end{tabular}
\end{scriptsize}
\end{center}
\end{table*}

\begin{table*}
\begin{center}
\contcaption{Spectral parameters and AGN contribution for the 71 ULIRGs in our sample.}
\begin{scriptsize}
\begin{tabular}{lcccccccccc}
\hline \hline
Source & $\alpha_6$ & $\tau_6$ & $\alpha_\mathit{bol}$ & $\tau_\mathit{ice}$ & $\tau_\mathit{hac,1}$ & 
$\tau_\mathit{hac,2}$ & $\alpha_6$* & $\tau_6$* & $\alpha_\mathit{bol}$* & $V/L/X$ \\
\hline
IRAS 20100$-$4156 & 86$\pm$1 & 0.47$\pm$0.02 & 22$\pm$4 & $<2.44$ & 0.77$\pm$0.05 & 
0.67$\pm$0.03 & 92$\pm$1 & 1.00$\pm$0.02 & 24$^{+5}_{-3}$ & \small $\star^3 \star^5 \odot^{12}$ \\
IRAS 20414$-$1651 & $<2.2$ & $<0.07$ & $<0.1$ & $-$ & $-$ & $-$ & $<2.7$ & $<0.05$ & $<0.08$ & 
\small $\star^1 \star^4 -$ \\
IRAS 20551$-$4250 & 90$\pm$1 & 1.19$\pm$0.01 & 28$^{+5}_{-4}$ & 0.38$\pm$0.02 & 0.35$\pm$0.01 & 
$<0.09$: & 94$\pm$1 & 1.71$\pm$0.01 & 31$^{+5}_{-4}$ & \small $\otimes^3 \bullet^5 \bullet^{12}$ \\
IRAS 21208$-$0519 & $<0.9$ & $-$ & $<0.04$ & $-$ & $-$ & $-$ & $<0.8$ & $-$ & $<0.03$ & 
\small $\star^1 \star^4 -$ \\
IRAS 21219$-$1757 & $>95$ & $<0.01$ & 85$^{+11}_{-42}$ & $<0.01$ & $<0.01$ & $<0.01$ & 
$>98$ & 0.37$\pm$0.01 & 65$\pm$5 & \small $\bullet^1 \bullet^4 -$ \\
IRAS 21329$-$2346 & 43$\pm$2 & $<0.07$ & 3.1$^{+1.0}_{-0.7}$ & $<1.08$ & $<0.01$ & 
$<0.01$ & 55$\pm$2 & 0.50$\pm$0.05 & 3.1$^{+1.0}_{-0.7}$ & \small $\otimes^1 \odot^4 -$ \\
IRAS 22206$-$2715 & $<9.3$ & $<0.17$ & $<0.5$ & $-$ & $-$ & $-$ & $<11$ & $<0.15$ & $<0.4$ & 
\small $\star^1 - -$ \\
IRAS 22491$-$1808 & $<1.4$ & $-$ & $<0.07$ & $-$ & $-$ & $-$ & $<1.3$ & $-$ & $<0.04$ & 
\small $\star^1 - \star^{12}$ \\
IRAS 23128$-$5919 & 48$\pm$1 & 0.36$\pm$0.03 & 3.9$^{+1.1}_{-0.8}$ & $<0.01$ & $<0.01$ & 
$<0.01$ & 61$\pm$1 & 0.87$\pm$0.03 & 4.3$^{+1.2}_{-0.8}$ & \small $\odot^3 \bullet^5 \odot^{12}$ \\
IRAS 23234+0946 & 30$^{+2}_{-3}$ & $<0.05$ & 1.8$^{+0.6}_{-0.5}$ & 0.31$^{+0.07}_{-0.18}$ & 
$<0.01$ & $<0.01$ & 41$\pm$2 & 0.47$\pm$0.08 & 1.9$^{+0.7}_{-0.5}$ & \small $\otimes^1 \star^4 -$ \\
IRAS 23327+2913 & 73$\pm$1 & 0.86$\pm$0.03 & 11$\pm$2 & 0.11$\pm$0.08 & $<0.01$ & $<0.26$: & 
82$\pm$1 & 1.37$\pm$0.03 & 11$^{+3}_{-2}$ & \small $\otimes^1 \star^4 -$ \\
MRK 231 & 93$\pm$1 & $<0.12$ & 36$^{+5}_{-4}$ & $<0.17$ & 0.08$\pm$0.01 & 0.03$\pm$0.01 & 
97$\pm$1 & 0.68$\pm$0.01 & 51$^{+6}_{-5}$ & \small $\bullet^2 \bullet^4 \bullet^{12}$ \\
MRK 273 & 67$^{+1}_{-4}$ & $<0.01$ & 8.1$^{+1.9}_{-2.3}$ & 0.55$^{+0.01}_{-0.14}$ & 
0.49$^{+0.02}_{-0.09}$ & 0.17$^{+0.01}_{-0.03}$ & 76$\pm$1 & 0.45$\pm$0.01 & 8.4$^{+2.0}_{-1.4}$ & 
\small $\bullet^1 \bullet^4 \bullet^{13}$ \\
NGC 6240 & 65$^{+6}_{-8}$ & 0.64$^{+0.20}_{-0.24}$ & 7.5$^{+4.0}_{-3.0}$ & $<0.01$ & $<0.01$ & 
$<0.01$ & 76$^{+3}_{-6}$ & 1.15$\pm$0.18& 8.1$^{+3.6}_{-2.8}$ & \small $\otimes^2 \bullet^6 \bullet^{14}$ \\
4C +12.50 & 97$\pm$1 & 0.24$\pm$0.02 & 62$\pm$7 & 0.11$\pm$0.04 & $<0.01$ & 0.05$\pm$0.01 & 
98$\pm$1 & 0.74$\pm$0.02 & 63$\pm$7 & \small $\bullet^1 \bullet^4 \bullet^{11}$ \\
UGC 5101$^i$ & 81$^{+1}_{-3}$ & $<0.02$ & 16$\pm$3 & $<1.06$ & $<0.99$ & $<0.90$ & 83$\pm$1 & 
0.19$\pm$0.01 & 13$^{+3}_{-2}$ & \small $\otimes^2 \bullet^7 \bullet^{15}$ \\
\hline
\end{tabular}
\end{scriptsize}
\end{center}
\begin{flushleft}
\textit{References.} $^1$~Veilleux et al.~(1999a), $^2$~Veilleux et al.~(1995), 
$^3$~Duc et al.~(1997), $^4$~Imanishi, Dudley \& Maloney~(2006), $^5$~Risaliti et al.~(2006b), 
$^6$~Risaliti et al.~(2006a), $^7$~Imanishi, Dudley \& Maloney~(2001), $^8$~Iwasawa et al.~(2005), 
$^9$~Teng et al.~(2008), $^{10}$~Severgnini et al.~(2001), $^{11}$~Imanishi \& 
Terashima~(2004), $^{12}$~Franceschini et al.~(2003), $^{13}$~Balestra et al.~(2005), 
$^{14}$~Vignati et al.~(1999), $^{15}$~Imanishi et al.~(2003). \\
\textit{Notes.} $^a$~The value of $\tau_6$ is obtained allowing $\Gamma_\mathit{agn}=0.78$. 
$^b$~This is indeed considered an AGN detection. $^c$~A comparable minimum in the parameter space 
suggests a possible AGN component with $\tau_6=0.24$ and $\alpha_\mathit{bol}=1.1$ (per cent). 
$^d$~The value of $\tau_6$ is obtained allowing $\Gamma_\mathit{agn}=0.58$. $^e$~The value of 
$\tau_6$ is obtained allowing $\Gamma_\mathit{agn}=1.31$. $^f$~A comparable minimum in the parameter 
space suggests that this source could also be SB-dominated. $^g$~Due to the dispersion around the 
templates, this value of $\alpha_\mathit{bol}$ is too small to include this source among the safe 
AGN detections. $^h$~The value of $\tau_6$ is obtained allowing $\Gamma_\mathit{agn}=0.52$. 
$^i$~The value of $\tau_6$ is obtained allowing $\Gamma_\mathit{agn}=1.02$.
\end{flushleft}
\end{table*}
In the light of the elements collected so far, we are able to address again the 
question about the existence of extremely obscured AGN components that can not be detected 
even at 5--8~$\mu$m, but only contribute to the IR emission at longer wavelengths (e.g. 
at $\sim$25--60~$\mu$m). Such components would pull downward our estimate of 
$\log R^\mathit{sb}=-1.91$. An empirical measure of this quantity can be 
performed at lower IR luminosities on the B06 starburst sample, yielding an average 
of $\log R^\mathit{sb}=-1.66$. This latter value anyway requires some caution 
for two opposite reasons: some objects in B06 (e.g. NGC~1365, NGC~4945) show significant 
nuclear activity; on the other hand, we are dealing with nearby sources and aperture losses 
in the SL spectral orders can occur\footnote{We have adopted the same strategy of 
B06, scaling up all the spectral orders to match the LL1 flux density, and then 
normalizing to IRAS at 25~$\mu$m. By comparing the final spectra to the 12~$\mu$m 
IRAS fluxes we evince that a residual loss may be present, and provide an upper limit 
for the average of $\log R^\mathit{sb} \approx -1.5$.}. Disregarding such  
complications (these systematic effects are believed to compensate for each other), it turns 
out that the two distributions of $R^\mathit{sb}$ fairly overlap, due to their large 
dispersion. Even if they appear to be somewhat displaced with respect to each other, the 
variation is too small to support the hypothesis of a completely buried AGN population 
escaping this 5--8~$\mu$m spectral probe (or any other \textit{Spitzer}-IRS probe; see 
also Veilleux et al. 2009). This difference can be possibly ascribed to the overall physical 
properties of the starburst process itself; for example, in the ULIRG luminosity range the 
starburst is triggered by a major merger (e.g. Dasyra et al. 2006), while at lower luminosities 
it mostly takes place in isolated systems.

\section{Comparison with other diagnostics}

Our 5--8~$\mu$m AGN/SB decomposition has provided 50 convincing AGN detections out 
of 71 sources in our sample, confirming all the already known AGN components and uncovering 
some more. Conversely, the AGN detection rate among ULIRGs is $\sim$30 per cent at optical 
wavelengths (Veilleux et al. 1999a), and $<50$ per cent according to mid-IR high-ionization 
lines (Farrah et al. 2007). X-ray diagnostics is actually extremely powerful, but can be 
applied on such a large scale only collecting a huge integration time: after Teng et al. 
2005, we estimate that an average exposure of $\sim$30~ks per source is required in order 
to reach up to our detection rate with the present X-ray facilities. 
In Section~\ref{rd} we have briefly summarized the multiwavelength picture of 
ULIRGs which is obtained from the most effective diagnostic methods used so far. 
We now discuss in more detail how our results compare with those of previous works, 
allowing us also to overcome some of their limitations. 

\subsection{Optical}

\begin{figure}
\includegraphics[width=8.5cm]{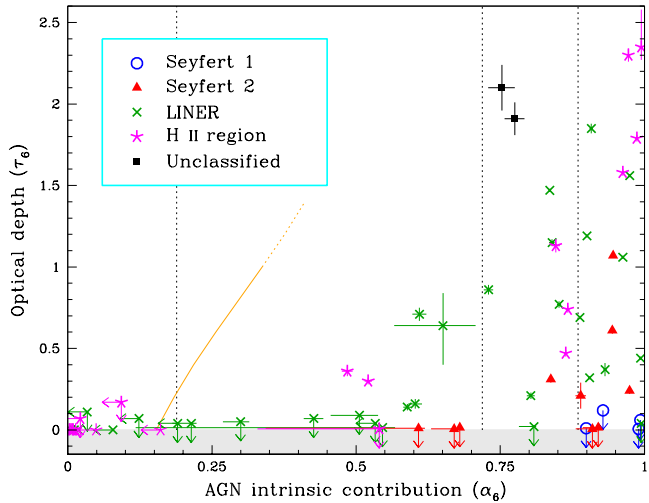}
\caption{Optical depth to the AGN component evaluated at 6~$\mu$m versus the AGN 
intrinsic contribution to the 5--8~$\mu$m emission. Each source is plotted with 
a different symbol according to its optical classification, as indicated in the 
box. The vertical \textit{black dotted lines} mark an AGN contribution to the bolometric luminosity 
of 1, 10 and 25 per cent, respectively. The crosswise \textit{orange curve} represents 
our empirical detection limit: due to the dispersion around our AGN/SB templates, an AGN 
component on the left of this line could be missed by our method (because its observed 
contribution is too faint). For the same reason, AGN detections close to this region of the 
plot are to be considered as tentative. The bolometric contribution of such components 
would be anyway negligible (a few per cent at most).} 
\label{opt}
\end{figure}

The nuclear regions of ULIRGs are affected by typical obscurations of several 
tens of magnitude at visible wavelengths, hence it is difficult to interpret 
the results of the diagnostics based on emission lines. None the less some ULIRGs 
exhibit in their spectra the broad optical recombination lines characteristic of 
Seyfert 1 galaxies. Even when a direct access to the broad line region is 
obstructed by larger amounts of dust, possibly aggregated in a toroidal shape 
as expected for type 2 Seyfert-like objects, the presence of a working AGN can 
be firmly established in the near-IR via the detection of broad permitted features 
or of the high-ionization 1.962~$\mu$m [Si~\textsc{vi}] line (Veilleux, Sanders \& Kim 1999). 
In principle, a hidden broad line region can also be looked for in polarized light, 
a technique that has been developed with success for nearby Seyfert 2 galaxies (after 
Antonucci \& Miller 1985) but is almost unavailable for ULIRGs (e.g. Pernechele 
et al. 2003) because of their larger distance, and possibly frustrated by the greater 
complexity in the geometrical structure of the dust (large covering factors and clumpiness 
are not easily reconcilable with the reflection scenario, nor with dichroic transmission).
In most cases only narrow features are observed, and it is necessary to resort to line 
ratios in order to classify a ULIRG. The separation of active nuclei and 
star-forming galaxies is based on empirical boundaries, discriminating the 
effects of photoionization by hot young stars from those due to non-thermal 
sources with a power-law continuum. This is in itself a limitation to the 
accuracy of this diagnostic method, which is further degraded by the extinction effects. 
According to Veilleux \& Osterbrock (1987), three equivalent diagnostic diagrams 
can be set up using the line ratios [O~\textsc{i}]~$\lambda$6300/H$\alpha$, 
[N~\textsc{ii}]~$\lambda$6583/H$\alpha$ and 
[S~\textsc{ii}]~($\lambda$6716+$\lambda$6731)/H$\alpha$ along with 
[O~\textsc{iii}]~$\lambda$5007/H$\beta$ (the latter distinguishes between 
a type 2 Seyfert galaxy and a LINER). It turns out that a significant number 
of ULIRGs do not belong to the same spectral type in all these diagrams, with 
some degree of overlap among LINERs and H~\textsc{ii} regions (Veilleux et al. 1999a). 
In any case, no evidence of AGN activity is clearly found within these two classes. \\
In Fig.\ref{opt} the AGN intrinsic contribution to the 5--8~$\mu$m emission and 
its obscuration, as obtained with our technique, are plotted together with the optical 
properties. With some scatter, the location of each source correlates with its optical 
classification. In particular, there is a tight agreement for Seyfert-like ULIRGs: type 1 
objects are found in the bottom right-hand corner, with a large AGN content and vanishing 
obscuration, while the average type 2 hosts a slightly dimmer and obscured AGN. 
The interpretation of the plot is less straightforward when one considers also the sources that 
are classified as H~\textsc{ii} regions or LINERs. It is clear that optical 
spectroscopy, besides giving incomplete information about the relative AGN/SB 
contribution, can also be misleading: this is indicated by the presence of only one
type 2 Seyfert-like source among the nine entries with $\alpha_\mathit{bol}>0.25$ 
and $\tau_6>1$. Also four H~\textsc{ii} galaxies fall into this group, and they are 
all true AGN detections at a simple visual inspection, sharing a reddened and absorbed 
continuum and strongly suppressed aromatic features. These targets are hence perfect 
candidates for follow-up observations at X-ray wavelengths. As opposed to the unexpected 
scatter of H~\textsc{ii} galaxies, the wide dispersion of LINERs in Fig.\ref{opt}
confirms again that such systems are rather heterogeneous with respect to the nature of their 
energy source. Out of 33 LINERs comprised in our sample, 23 contain an AGN component, 
with a wide range of bolometric contributions. This fraction ($\sim$70 per cent) is 
very close to the detection rate achieved by collecting all multiwavelength diagnostics, but 
is significantly higher than those obtained from radio, optical and X-ray observations 
separately. Our decomposition method is then capable to resolve most of the ambiguity 
concerning the ionization mechanism in LINERs (see also Dudik, Satyapal \& Marcu 2009; 
Gonzalez-Martin et al. 2009). 

\subsection{Mid-IR}

A wealth of studies at mid-IR wavelengths have been published since the launch of 
the Infrared Space Observatory (ISO; Kessler et al. 1996), most of which rely on 
different mixtures of three basic diagnostics, i.e. hot dust continuum, aromatic 
emission and silicate absorption. The usual strategy makes use of single indicators 
such as the equivalent width of the 6.2 or 7.7~$\mu$m PAH feature, flux ratios 
involving fine-structure lines, the 9.7~$\mu$m silicate optical depth, the 
15 to 6~$\mu$m continuum colour. Pairs of these diagnostics are used 
to construct bidimensional diagrams in which the location of ULIRGs 
is compared with those of standard AGN and SB control samples (e.g. Genzel et 
al. 1998). Alternatively, one can attempt at reproducing the whole observed 
emission within a certain wavelength range by means of a limited number of 
components. Both spline continua plus variable PAH emission and fixed spectral 
templates have been employed. Our method follows the latter approach, and is 
qualitatively similar to those presented in Laurent et al. (2000) and Tran et 
al. (2001), with the great improvement in data quality made available by 
\textit{Spitzer}. It is difficult to find a quantitative correspondence among 
the results obtained on the individual objects, since these earlier works were 
intended primarily to derive the statistical properties of the ULIRG population 
as a whole. 
\begin{figure}
\includegraphics[width=8.5cm]{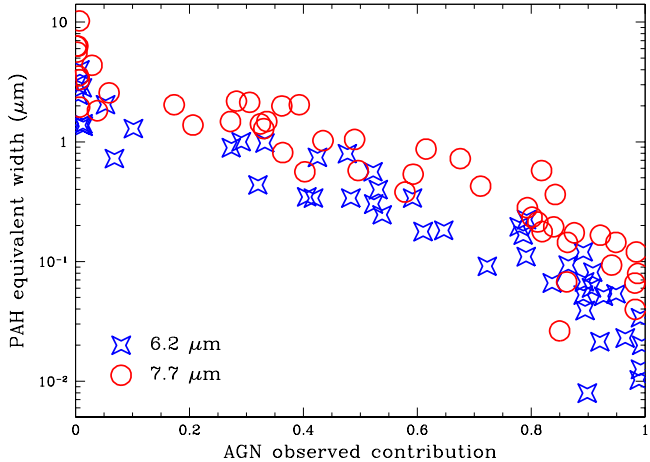}
\caption{Equivalent widths of the 6.2~$\mu$m (\textit{blue diamonds}) and the 
7.7~$\mu$m (\textit{red rings}) PAH features (from Veilleux et al. 2009) plotted 
against our estimate of the AGN observed contribution to the 5--8 $\mu$m continuum, 
evaluated at the same wavelength. The typical error bars are comparable to 
the symbol size.}
\label{mid}
\end{figure}
A possible example of a more specific comparison is illustrated in Fig.\ref{mid}, 
where the AGN observed contribution to the 5--8~$\mu$m emission of 48 
sources is compared to the equivalent widths of the 6.2 and 7.7~$\mu$m aromatic features 
(from Veilleux et al. 2009). The expected anti-correlation between these quantities 
can be appreciated, and the scatter is highly reduced with respect to the measurements 
performed before the \textit{Spitzer} era (e.g. Rigopoulou et al. 1999). At the same 
time it is clear that reddening extensively shapes the spectra of ULIRGs, and its effects 
have a substantial impact when estimating the AGN/SB bolometric contribution. \\
Another way to probe the degree of absorption and the geometrical structure of the dust, 
and test the characterization we offer of these issues, is provided by the silicate 
feature at 9.7~$\mu$m: its opacity, together with the intensity of aromatic emission, 
can be plotted in a diagram useful not only for working as a classification scheme, 
but also for tracing the possible evolutionary path of a source. The silicate feature 
has not been studied in the present work. We anyway point out the good agreement between 
our findings about the most notable sources in our sample and their position in such a 
diagnostic diagram (Spoon et al. 2007). In particular, the silicate strength appears to 
have a positive correlation with the reddening we have estimated for the AGN component. 

\subsection{High-ionization lines}

As mentioned before in relation with the 1.962~$\mu$m [Si~\textsc{vi}], the 
presence of fine-structure lines from highly-ionized atoms has a powerful diagnostic 
value in distinguishing between black hole accretion and star formation in ULIRGs.  
Also in the mid-IR, the mere detection of features such as the [Ne~\textsc{v}] 
$\lambda$14.32 line is the clear mark of an active nucleus, because of the required 
ionization energy of 97.1~eV. Again, by combining suitable line ratios among themselves 
or with different AGN/SB indicators, useful diagnostic diagrams can be set up, 
that make it possible to point out the main average trends but can be 
misleading if applied to single objects. \\
Recently Farrah et al. (2007) have analyzed the high-resolution \textit{Spitzer}-IRS 
spectra of 53 local ULIRGs, 27 of which are also included in our sample. This 
allows a meaningful comparison between our AGN/SB decomposition method and 
diagnostics based on fine-structure lines. Focusing on the common sources, 
three of them have been used for the construction of our SB template, while 
signatures of AGN activity have been found in each of the remaining 24 according 
both to Paper~I and the present work. We therefore look for a 
confirmation of this activity through high-ionization lines. The simplest 
approach is of course the detection of individual features. [Ne~\textsc{v}] 
$\lambda$14.32 is found in ten objects, in nine of which we detect an AGN 
component as well. The only exception is IRAS~20414$-$1651, that we have 
assumed to be a prototypal starburst because of its prominent PAH features 
and cold mid-IR colours. IRAS~20414$-$1651 is actually the only case without 
the simultaneous detection of [O~\textsc{iv}] $\lambda$25.89, another usual 
clue of hard radiation fields. On the contrary, IRAS~10378+1109 tentatively shows the 
oxygen line but no neon emission\footnote{This detection is not actually confirmed by 
Veilleux et al. (2009), as well as the detections in 4C +12.50; the [Ne~\textsc{v}] $\blambda$24.32 
line is instead found in IRAS~15206+3342.}. It is not obvious to relate the absence of [Ne~\textsc{v}] 
$\lambda$14.32 to a higher degree of obscuration. Both MRK~231 and IRAS~15462$-$0450, 
for instance, are not detected in [Ne~\textsc{v}] $\lambda$14.32 despite their nature 
of type 1 Seyfert galaxies at optical wavelengths. The same holds for IRAS~07598+6508. 
None the less, it is worth noting that the median value of the optical depth to the AGN 
component is $\sim$0.3 for the nine ULIRGs with a [Ne~\textsc{v}] $\lambda$14.32 detection 
and $\sim$0.5 for the remaining 15 composite sources, and indeed the argument of a larger 
obscuration seems to be reasonable for many of them. A high covering factor of the 
absorbing dust can even prevent the formation of a narrow line region and the 
production of fine-structure lines. A complete revision of both line fluxes and 
upper limits can be found in the recent work of Veilleux et al. (2009), that also 
includes 25 additional sources within our sample. The considerations made so far are widely 
corroborated. There are six more objects with a [Ne~\textsc{v}] $\lambda$14.32 line: only one 
(IRAS~13335$-$2612) is classified as SB-dominated after our 5--8~$\mu$m spectral analysis. 
Also the loose correlation between successful detections and little opacities is possibly 
emphasized. \\ 
The line ratios allow more complete diagnostics in combination with some other AGN/SB 
indicator over the electromagnetic spectrum. All the diagrams proposed by Farrah et 
al. (2007) point to star formation as the dominant power supply in ULIRGs: the active 
nucleus is as powerful as the starburst in only $\sim$20 per cent of the sources, and 
of lesser significance in more than half. These fractions are in perfect agreement with 
our estimates. The effectiveness of mid-IR fine-structure lines in ULIRG diagnostics is 
anyway restricted to the finding of general trends, since no firm quantitative constraint 
can be put on the AGN bolometric contribution (except for some remarkable examples), and 
even qualitative considerations may lead to mistaken conclusions on individual objects. 

\subsection{Hard X-rays}

Since ULIRGs are faint X-ray emitters, with a 2--10~keV continuum flux ranging from 
$\sim$10$^{-5}$ to a few $\times$10$^{-3}$ of their overall IR output, only a limited number 
of them has been observed successfully with the \textit{Chandra} and \textit{XMM-Newton} 
satellites. We can therefore provide only a brief compilation of the main results that have been obtained 
so far. Clear X-ray signatures of AGN activity are a 2--10~keV luminosity exceeding 
10$^{42}$~erg~s$^{-1}$ and a large column density ($N_H > 10^{22}$~cm$^{-2}$) absorbing 
an unresolved component. The detection of the iron K$\alpha$ fluorescent line at 6.4~keV 
with an equivalent width of $\sim$1~keV also indicates the presence of an obscured active 
nucleus, whose primary radiation is reflected by a \textit{cold} mirror of gaseous material 
(for a detailed review about this subject see Reynolds \& Nowak 2003). 
All these indicators have been found in IRAS~19254$-$7245 (Braito et al. 2003, 2009). 
MRK~231 lacks a strong iron line, but the direct AGN emission emerges above 20~keV 
(Braito et al. 2004); moreover this source exhibits significant nuclear variability 
on timescales of a few hours (Gallagher et al. 2002). Both nuclei of NGC~6240 have 
hard X-ray emission that possibly dominates the global energetics, with heavy absorption 
and a prominent neutral iron line (Vignati et al. 1999; Komossa et al. 
2003). IRAS~05189$-$2524 is a typical Compton-thin Seyfert 2 galaxy, in which the estimated 
luminosity of the AGN component can not account for the huge IR emission (Severgnini et 
al. 2001); also MRK~273 (Balestra et al. 2005) and UGC~5101 (Imanishi et al. 2003) 
are likely SB-dominated, although an AGN is detected above $\sim$3--4~keV. The hard X-ray 
luminosity and spectral properties suggest an AGN contribution to IRAS~20551$-$4250 
and, tentatively, to IRAS~23128$-$5919 and IRAS~20100$-$4156, while no convincing 
evidence of AGN is found in IRAS~12112+0305, IRAS~14348$-$1447, IRAS~15250+3609, 
IRAS~17208$-$0014 and IRAS~22491$-$1808 (Franceschini et al. 2003; Ptak et al. 2003). \\
Even when measured, the luminosity of the above-mentioned AGN components generally lies 
in the range of Seyfert galaxies. In order to represent the major energy source of the 
ULIRG  bolometric luminosity they should be located behind a column density of obscuring 
material approaching 10$^{25}$~cm$^{-2}$. It is still matter of debate if this is the 
case for ARP~220, the nearest ULIRG, inside which an elusive active nucleus is suspected 
to be at work. A strong iron line has indeed been detected in ARP~220 by Iwasawa et al. 
(2005), challenging the origin of the 2--10~keV emission from X-ray binaries and hence the 
pure starburst interpretation (for the X-ray emission processes in starburst galaxies 
see Persic \& Rephaeli 2002). \\ 
A handful of additional sources has been recently studied by Sani, Risaliti 
\& Salvati (in preparation) using also the \textit{BeppoSAX} observations up 
to 100~keV. The most striking result concerns IRAS~12072$-$0444, which is 
detected only at energies beyond 20~keV and shows up as a Compton-thick source 
harbouring a type 2 quasar with a column density $N_H \simeq$4$\times 10^{24}$~cm$^{-2}$. 
IRAS~08572+3915 is not detected but is suggested to be a type 2 quasar as well, 
whose reflection efficiency is very low by virtue of a cocoon-like obscuring geometry. 
Finally, on the basis of the ratio between the 2--10~keV and the IR emission, 
IRAS~10190+1322 and IRAS~10494+4424 confirm their SB-dominated nature, while the 
presence of an AGN component can not be ruled out in IRAS~00188$-$0856, IRAS~01003$-$2238 
and IRAS~16090$-$0139: according to their \textit{Spitzer}-IRS spectra and the 
findings of the present work these latter objects are believed to host a buried 
AGN, and are expected to be absorbed even at 10~keV. Regrettably one of them 
(IRAS~16090$-$0139) was observed but not detected by \textit{BeppoSAX}, and 
the upper limit to its flux above 20~keV is not constraining; the other two 
have not been observed at all. \\ 
In conclusion, our results are in qualitative agreement with all the pieces of 
evidence from X-ray studies. The X-ray diagnostics of ULIRGs is very powerful, 
yet our 5--8~$\mu$m analysis provides fully consistent indications and, at present, 
gives access to a much larger number of sources.

\section{Discussion}
\label{dis}

Having illustrated the global results of our study and their consistence 
with the multiwavelength picture of ULIRGs, in this Section we discuss 
in quantitative terms the reliability of alternative AGN templates. 
We also address some uncertain or anomalous cases, and their consequences 
with respect to the dust properties within the ULIRG class. We finally 
investigate whether our analysis is able to recover the claimed correlation 
between AGN activity and bolometric luminosity. 

\subsection{Fitting of peculiar sources}

\begin{figure}
\includegraphics[width=8.5cm]{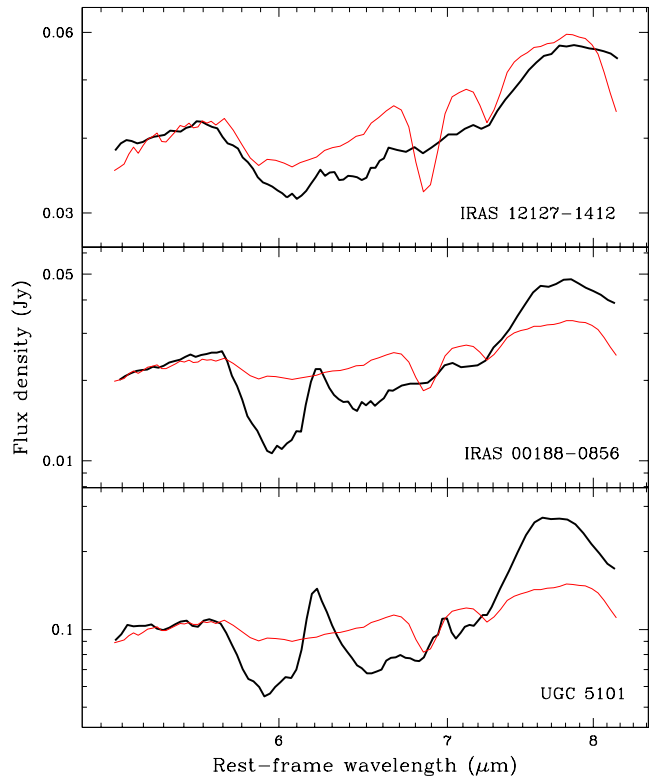}
\caption{The puzzling spectral trend of the \textit{anomalous} sources in 
our sample, compared to the 5--8~$\mu$m emission of IRAS F00183$-$7111 
(\textit{thin red line}, scaled to match the putative hot dust continuum). 
The apparent flatness can indeed be connected to extreme absorption of the 
AGN component within a different extinction scenario.} 
\label{odd}
\end{figure}

In our model the great diversity of the observed ULIRG spectra 
can be entirely ascribed to the relative amount of the 
AGN contribution and its obscuration, intended as both absorption and 
reddening. The main features are adequately reproduced 
and a good fit is obtained for most of the spectra in our sample, with 
only a few exceptions whose analysis required additional assumptions. 
The main issue is the possible flattening of the AGN component (as found by N07), 
which entails a relaxation of our general template. The representative case 
is that of IRAS~07598+6508, which is unambiguously an AGN-dominated 
source: its continuum trend can be reproduced by simply allowing the 
AGN spectral index to be $\sim$0.6. IRAS~19254$-$7245 (the \textit{Superantennae}) 
is instead a composite system with regular absorption features. In order 
to obtain a good fit we need again to provide a flatter AGN component: 
including its intrinsic slope among the degrees of freedom we obtain 
a value of $\sim$0.5, with a 6~$\mu$m optical depth of $\sim$0.2. \\
The remaining three particular cases (that in the following will also be 
labelled as \textit{anomalous}) are more interesting, in that the 
supplementary presence of a broad and deep absorption over most of the 
5--8~$\mu$m range causes the failure of our continuum-based AGN/SB 
decomposition. All these objects are likely to host a completely buried 
AGN; their emission is shown in Fig.\ref{odd}, where they are compared to 
the benchmark spectrum of IRAS~F00183$-$7111 (Spoon et al. 2004; Nandra 
\& Iwasawa 2007). Qualitatively, IRAS~12127$-$1412 is the most striking 
object in our sample, since the broad absorption completely replaces 
the 6.2~$\mu$m PAH emission (no trace of which is clearly visible) and 
stretches longward of 7~$\mu$m. In IRAS~00188$-$0856 and UGC~5101 a strong,
hard spectrum (closely resembling the hot dust continuum of AGN) 
stops abruptly: this occurs in some other sources as well, but the 
putative AGN continuum is rather steep and only a proper modelling 
of the absorption profiles is entailed. On the contrary, the three 
cases at issue suggest once again a flattening of the AGN component 
with respect to $\Gamma_\mathit{agn}$=1.5, and the aromatic features, 
if present, seem to emerge from the bottom of a deep trough. 
A customized fitting procedure is therefore needed. The observed 
continuum is assumed to be AGN-dominated at $\sim$5.5~$\mu$m, and 
since the available window is very narrow we have employed the 
15~$\mu$m flux as a pivotal point of the fit. The AGN intrinsic 
slopes obtained in such way range from $\sim$0.8 to $\sim$1.3, 
with $\tau_6$ between $\simeq$0.4 and $\simeq$0. We have then 
frozen the continuum and included the SB component and the absorption 
profiles in the next fitting step. The silicate feature 
is also required longward of $\sim$7.5~$\mu$m, as in other few 
objects. The physical interpretation of these \textit{anomalous} 
spectra, concerning both flatness and extreme absorption, will be 
discussed in detail in the following subsections. 

\subsection{The AGN intrinsic spectral slope}

\begin{figure}
\includegraphics[width=8.5cm]{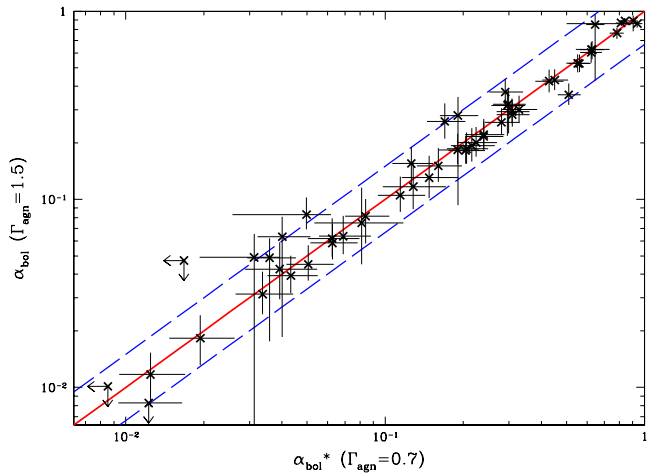}
\caption{Different estimates of the AGN bolometric contribution ($\alpha_\mathit{bol}$ and 
$\alpha_\mathit{bol}$*) as a function of its intrinsic spectral slope. The vertical axis 
shows the values obtained for $\Gamma_\mathit{agn}$=1.5, the horizontal one those obtained 
for $\Gamma_\mathit{agn}$=0.7. The \textit{red solid line} marks the equality between the 
two estimates, while the \textit{blue dashed lines} indicate a deviation of 50 per cent: 
each entry falls inside this region. For clarity, 17 sources that are confirmed to 
be SB-dominated in both cases have been omitted, and fall outside the region plotted above. 
We recall that, due to the dispersion around our templates, safe AGN detections require 
$\alpha_\mathit{bol}\ga$0.01.} 
\label{a2a}
\end{figure}

A power-law behaviour of the 5--8~$\mu$m hot dust continuum in AGN is confirmed 
by the observations, which however provide different values of the spectral index. 
We recall that the slope of the AGN template can not be treated as 
a free parameter, because of its degeneracy with the optical depth. 
Our assumption of a steep intrinsic gradient has been explained in 
Section~\ref{sd}. Conversely, we have just presented a few peculiar 
cases whose spectra are better reproduced by a lower value of 
$\Gamma_\mathit{agn}$. It is then worth testing how our estimates 
of the AGN bolometric contribution depend on this choice. \\
Since the quality of our fits is good and the SB template is not modified, we 
expect the observed AGN component to be nearly invariant. Hence the flatter the 
intrinsic continuum, the larger the optical depth, and consequently the AGN flux 
amplitude. As a first approximation, the relative \textit{weight} of the AGN component 
is then increased after the flattening of the intrinsic spectral shape. We have 
performed again the fitting of all our sources, modifying the AGN template and 
assuming the new value of $\Gamma_\mathit{agn}$=0.7 for its slope, as suggested 
by N07. The goodness of our fits remains virtually unchanged, and no convincing 
additional AGN detection is found. As forecast, however, both the optical depth 
and the AGN amplitude increase remarkably, because of the degeneracy between 
$\Gamma_\mathit{agn}$ and $\tau_6$ connected to the assumed \textit{recipe} 
for the extinction law. The systematically larger AGN flux also affects our 
estimate of $R^\mathit{agn}$, from the fitting of the $R$--$\alpha_6$ relation 
of Eq.(\ref{ra}). The modified values of the 6~$\mu$m to bolometric ratios are 
the following:
\begin{displaymath}
\log R^\mathit{agn}=-0.36^{+0.06}_{-0.07} \hspace*{10pt} \textrm{and} \hspace*{10pt} \log R^\mathit{sb}=-1.91^{+0.02}_{-0.02}.
\end{displaymath}
With respect to the estimates of Section~\ref{ae}, $R^\mathit{sb}$ is confirmed (the SB 
template is not changed), while $R^\mathit{agn}$ turns out to be $\sim$0.2~dex higher. 
The revised values of $\alpha_\mathit{bol}$ are again computed from Eq.(\ref{ab}), 
where now $R^\mathit{agn}/R^\mathit{sb}\simeq$35. A striking result is obtained 
as a consequence of the correlation between $\alpha_6$ and $R^\mathit{agn}$, i.e. 
a moderate variation in the slope of the AGN template has no significant effect 
on $\alpha_\mathit{bol}$. This invariance can be better appreciated in Fig.\ref{a2a}, 
while Tab.\ref{t3} lists the two alternate values of both the AGN bolometric contribution 
and the optical depth. 
This is of course a notable finding. Nevertheless we emphasize once again that our original 
template appears to be more suitable to describe the AGN component inside a ULIRG, in spite 
of the peculiar cases. The alternative scenario, in fact, implies that virtually all the 
detected AGN components are heavily reddened; as a consequence, the distribution of continuum 
opacities is discontinuous and strongly biased towards the highest values. This anomaly 
can not be explained even allowing a far larger dispersion in the 5--8~$\mu$m AGN spectral 
shapes than that observed by N07. 

\subsection{Absorption features}

We have already mentioned how the SB component is affected by internal extinction only; 
the large spread in the optical depth of the 9.7 and 18~$\mu$m silicate absorption 
features within the SB class suggests a wide range of dust properties, a full investigation 
of which would require a larger spectral coverage and a more detailed model than those 
considered in this work. Concerning the AGN component, it is not possible to put constraints 
on the nature or geometry of the cold screen blocking the view to the nuclear region.
However, we are able to discuss the properties of the main absorption features and the 
accuracy of a power-law extinction curve. \\
In general, the AGN and SB templates we have adopted are sufficient to account for 
all the major features observed in the 5--8~$\mu$m spectra of ULIRGs. Minor features, 
when involved, have been reproduced through gaussian profiles, but do not alter the 
AGN/SB decomposition. There are indeed plenty of possible absorption features falling 
in the 5--8~$\mu$m range, as can be also inferred from the observations of a highly-obscured 
environment like the Galactic Centre. Notably, none of these has been found 
in our spectra without the simultaneous detection of a significant AGN contribution 
to the continuum emission. A characteristic couple of features is frequently distinguished 
in composite sources, centred at $\sim$6.85 and $\sim$7.25~$\mu$m and ascribed 
to the C--H bending modes of aliphatic hydrocarbons (also known as hydrogenated amorphous 
carbons, HAC). They usually display a regular gaussian profile, and their relative depth 
can be used to constrain the chemical properties of the dust grains (see also 
Dartois \& Mu{\~n}oz-Caro 2007). The most sensitive point for our study is anyway 
the possible presence of another absorption feature, centred at $\sim$6.0~$\mu$m, 
that is commonly attributed to a mixture of ices and is much more complex.  
Although this profile can be strongly asymmetrical, with an elongated red wing, 
it is usually so shallow and/or regular that a rigorous selection of its shape is 
not required. In the mentioned case of Sgr~A* the 6~$\mu$m feature would be 
optimally fitted by means of pure H$_2$O ice at 30~K (Chiar et al. 2000), and 
this consequently represents a reasonable modelling for our purposes. We 
therefore decided to adopt the pure water ice profile at 30~K (provided by 
the Leiden Observatory database) as the backbone of the 6~$\mu$m absorption 
in our fits. \\ 
As discussed in the following, this assumption is challenged in only a few objects 
hosting a deeply enshrouded AGN (including the three \textit{anomalous} spectra, where 
it actually fails). The plain identification in terms of water ice, in fact, would 
lead to a large overestimate of the optical depth in the companion 3~$\mu$m absorption 
band. A simultaneous fit of both profiles in Sgr~A* drives only a partial explanation 
of the 6~$\mu$m feature as the product of a mixture of ices, i.e. H$_2$O:NH$_3$:CO$_2$ 
(100:30:6) at 15~K and pure HCOOH at 10~K (Chiar et al. 2000). Most of the residuals 
are accounted for by a 6.2~$\mu$m component whose origin is supposed to be aromatic. 
This remarkable example proves how the exact nature of the whole 6~$\mu$m 
feature is still poorly known. Yet the formation by surface chemistry of many different 
species in the icy mantles of the dust grains suggests the possibility of large 
variations in its profile, as a function of the chemical and physical properties 
of each source. This is indeed what we find out, since the central wavelength 
(actually the wavelength of minimum flux) appears to shift within the 
$\sim$5.9--6.2~$\mu$m range, and also the blue and red wings tend to be 
slightly different for the objects whose optical depth\footnote{All the 
optical depths concerning individual features refer to the AGN component only, 
and are evaluated considering the observed (i.e. not corrected for reddening) 
AGN continuum from the best fits.} (hereafter $\tau_\mathit{ice}$) is larger than 
that of Sgr~A*. A sizable value of $\tau_\mathit{ice}$ can be considered itself 
a qualitative indicator of a significant AGN contribution, but in extreme cases 
it can even prevent a straightforward AGN/SB decomposition. \\
In the light of the above considerations, a likely interpretation of the broad 
absorption band characterizing some of our 5--8~$\mu$m spectra is the blending of 
the 6~$\mu$m ice profile with the pair of HAC features around 7~$\mu$m. 
Evidence of a dramatic broadening of the 6.85~$\mu$m line is actually found 
in the observations of several embedded protostars (Keane et al. 2001). 
Moreover, according to a recent spectroscopic survey of young stellar objects, 
both water ice and HAC seem to play a deficient role among the carriers of the 
most enigmatic 5--8~$\mu$m depressions (Boogert et al. 2008). This could 
provide a physical explanation also to the spectral peculiarities observed in  
the three \textit{anomalous} cases, in which the deviations from the standard $\sim$6~$\mu$m 
and 6.85~$\mu$m profiles give rise to a formless band that undermines our approach. 

\subsection{Absorption versus reddening}

\begin{figure}
\includegraphics[width=8.5cm]{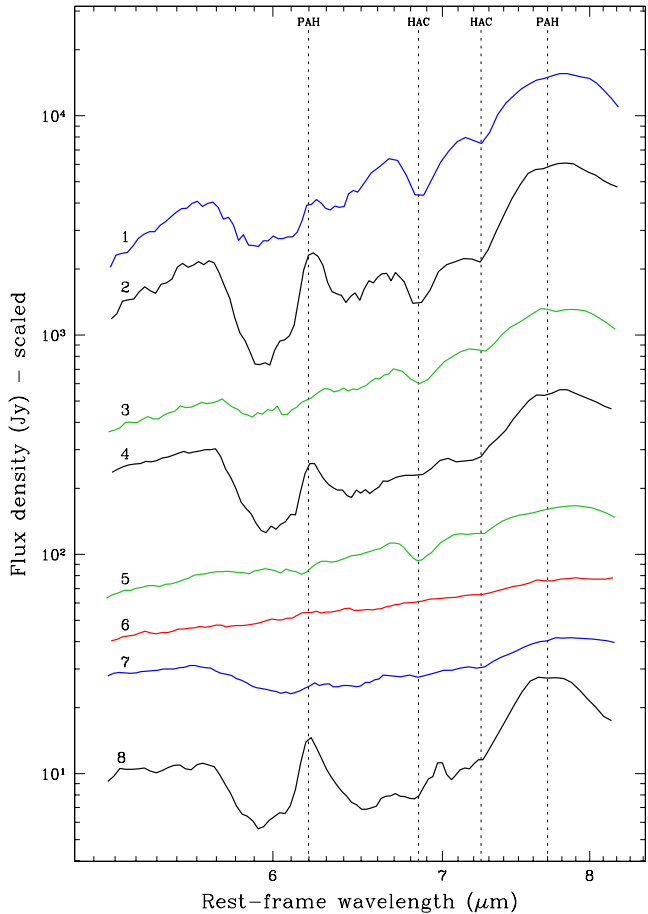}
\caption{From top top bottom, the 5--8~$\mu$m spectra of the following sources are 
plotted in a slope sequence with decreasing steepness: \textit{(1)}~IRAS~01298$-$0744, 
\textit{(2)}~IRAS~20100$-$4156, \textit{(3)}~IRAS~16156+0146,  \textit{(4)}~IRAS~00188$-$0856,  \textit{(5)}~IRAS~08572+3915, \textit{(6)}~IRAS~21219$-$1757,  \textit{(7)}~IRAS~12127$-$1412 
and \textit{(8)}~UGC~5101. Colours are used to denote the consequence of absorption 
(\textit{red}, \textit{green}, \textit{blue} and \textit{black} in order of growing 
significance). This contrast shows the wide variety in both reddening and absorption 
characteristic of ULIRGs.} 
\label{seq}
\end{figure}

\begin{figure}
\includegraphics[width=8.5cm]{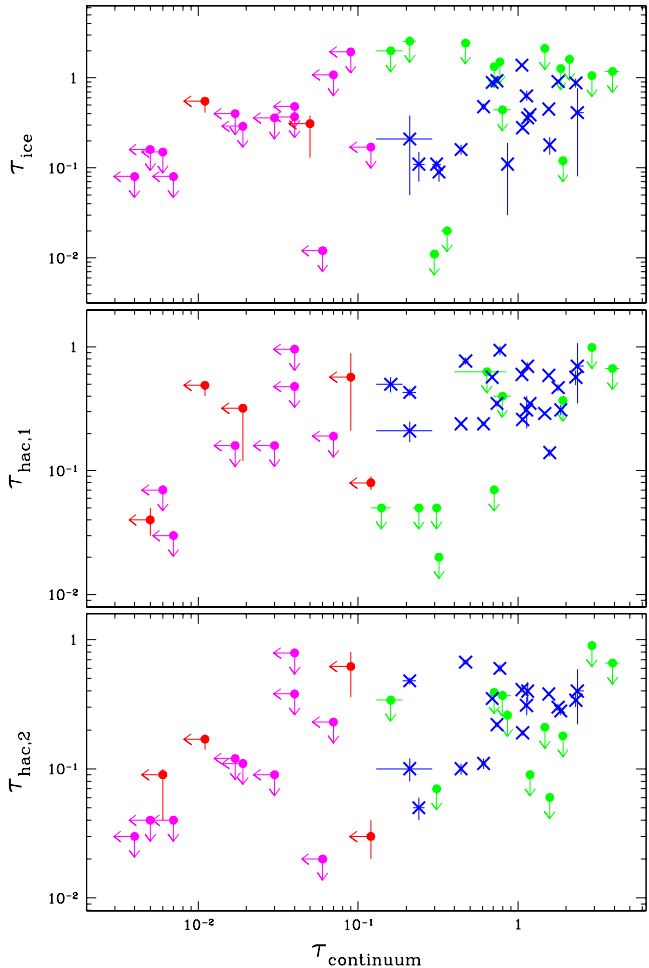}
\caption{Opacity of absorption features versus continuum opacity, inferred from 
the reddening of the AGN component. From top to bottom, the three panels show the 
qualitative correlation between the optical depth of the main observed features 
(ascribed to water ice at $\sim$6.0~$\mu$m and hydrogenated amorphous carbons at 
$\sim$6.85~$\mu$m and $\sim$7.25~$\mu$m) and the optical depth to the hot dust continuum. 
In each plot, the \textit{blue crosses} refer to well-established measurements of both 
quantities, while the other points render in a different colour code all the cases involving 
upper limits. Only the 50 AGN detections are considered, since no absorption is found 
in pure starbursts. In a few cases the observed AGN component is too faint and no sort 
of indication can be evinced about additional features: for this reason, the number of 
effective entries is 48, 43 and 43, respectively. The values of the optical depth for the 
three \textit{anomalous} sources are computed assuming a flat AGN intrinsic continuum and 
a quasi-grey extinction law (see the discussion in the text) and are provided as upper 
limits on the vertical axis, since the identification of the single features is unclear. 
If we take into account the secure entries only, we evince that no strict proportionality 
holds between reddening and absorption, nevertheless they positively correlate. This 
evidence is not controverted by the other points, whose constraints are very limited. 
We conclude that the appearance of broad absorption coupled with reddening deficit is 
confidentially due to a flat extinction law.} 
\label{abs}
\end{figure}

An accurate chemical analysis goes beyond the aims of the present work, anyway  
a further line of investigation can be pursued, involving the study of possible 
correlations between the optical depth of the most prominent absorption features 
and the reddening of the dust continuum. In Fig.\ref{seq} some representative 
ULIRG spectra have been arranged in a slope sequence with decreasing steepness 
with respect to the observed AGN component. Different colours have been used 
in order to flag the absorption degree and avoid ambiguities in the following 
considerations. Through a simple comparison among the adjacent spectra, it is 
clear that the intensity of the absorption features varies considerably within 
the same range of continuum slope. The effects of extreme absorption are evident 
e.g. in the \textit{black} spectra (IRAS~20100$-$4156, IRAS~00188$-$0856 and UGC~5101), 
all of which also display strong aromatic emission longward of 6~$\mu$m. At shorter 
wavelengths the observed flux density is unduly intense to share a common origin 
with the PAH features: this can be easily inferred by considering the shape of 
our SB template. If such emission is to be interpreted as the AGN-related hot 
dust continuum, one is in the presence of really contrasting signatures, since 
the aromatic nature of both the 6.2 and 7.7~$\mu$m peak is beyond dispute. 
Conversely, in cases like the \textit{blue} spectrum (IRAS~01298$-$0744) on 
the top of Fig.\ref{seq}, the possibility that the tentative 6.2~$\mu$m feature is 
actually a ridge in the absorbed continuum mimicking the PAH emission can not be ruled 
out. Moreover the supposed 7.7~$\mu$m companion appears to be broad and roundish, 
as usually observed in the AGN-dominated spectra (as also the \textit{green} ones, 
i.e. IRAS~16156+0146 and IRAS~08572+3915): this is rather a narrow window of 
unabsorbed continuum, that immediately declines into the silicate trough.
A possible explanation to the \textit{anomalous} spectra can be the coexistence 
of a buried AGN component of moderate luminosity with a powerful starburst. In this 
scenario the direct AGN emission is quenched by a dense screen, producing deep 
absorption bands whose carriers are unknown (perhaps various ices and HAC, see 
above). The sharp disappearance of the hot dust continuum is justified if the 
blending reaches up to the 9.7~$\mu$m silicate feature. A hint in this direction 
is provided by the comparison between the contiguous spectra of IRAS~21219$-$1757 
(\textit{red}, unabsorbed) and IRAS~12127$-$1412 (\textit{blue}): both these 
sources can be considered as pure AGN at 5--8~$\mu$m. In the former we observe 
a featureless and uniform continuum, which is a reasonable candidate to outline 
the true hot dust emission of the latter source as well. In this case, 
however, the measured 8~$\mu$m flux density is well below that expected from 
the extrapolation of the putative continuum slope. Hence it possibly represents 
only a saddle point between the broad $\sim$7 and 9.7~$\mu$m absorption bands. \\
The qualitative considerations above are of course insufficient to claim 
that very unusual physical conditions are responsible for the emergence of 
\textit{anomalous} spectral shapes and absorption/reddening polarities. 
In order to properly investigate these aspects and characterize the 
\textit{anomalous} sources, a detailed analysis is needed from \textit{K}-band up 
to $\sim$20~$\mu$m. This would make it possible to address the problems 
connected to the uncertain location of the continuum. For a few of our 
targets this spectral range is nearly covered in its entirety, and interesting 
indications come from \textit{L}- and \textit{M}-band studies in which strong aliphatic and 
CO absorption features are detected: the red wing of the 4.65~$\mu$m CO 
profile is indeed entering the very beginning of the \textit{Spitzer}-IRS 
window in several absorbed objects of our sample. None the less a joint 
modelling of the 2--20~$\mu$m SED is at present frustrated by the disparity 
in signal to noise and the scatter in flux calibration among the different 
ground-based and space facilities involved in the observations. \\
Summarizing, there seems to be no clear correlation between the degree 
of continuum reddening and the optical depth of individual absorption 
features. From a quantitative point of view this has already been established 
by Sani et al. (2008) in an extensive \textit{L}- and \textit{M}-band study of five bright ULIRGs. 
The presence of \textit{anomalous} sources in our sample challenges even the 
possibility of a step-wise correlation, which is instead found shortward 
of 5~$\mu$m: neither the detection of broad absorption features implies invariably 
a sizable reddening, nor a flat continuum is regularly unabsorbed. This should not 
be surprising, since ULIRGs have a very disturbed morphology and large variations 
in chemical composition and geometrical structure of the dust are naturally expected. 
Yet it is possible to reconcile the observation of significant absorption features 
alongside apparent lack of reddening without invoking special dust distributions and 
covering factors. In Fig.\ref{abs} the optical depths of the main absorption features 
are plotted against the optical depth to the hot dust continuum. All these values are 
also included in Tab.\ref{t3}. By focusing on the confident entries only, the positive 
correlation between absorption and reddening is tentatively recovered. This suggests 
that modifications of the extinction law are possibly involved. 

\subsection{The extinction law}

Conforming to our interpretation, both reddening and deep absorption are related to 
the presence of a compact power source, and must be due almost exclusively to cold 
material in the surroundings of the AGN component. We therefore argue that a reddened 
continuum is not observed in the \textit{anomalous} sources because the actual extinction 
law applicable to their AGN component has a softer wavelength dependence. Within the 
power-law prescription $\tau_\lambda \propto \lambda^{-1.75}$ that has been adopted 
after Draine (1989), the ratio between the values of the optical depth at the red and 
blue ends of our fitting region is $\tau_\mathit{red}/\tau_\mathit{blue}\simeq$0.44; 
this allows us to reproduce the steepest continuum gradients. Yet, according to the 
observations, the extinction curve can be safely described by a power law only up 
to $\sim$4--5~$\mu$m, and becomes rather controversial in the 5--8~$\mu$m range. 
For a long time, in absence of an adequate coverage of this spectral region with 
scientific data, the near-IR power-law trend was supposed to hold up to the 9.7~$\mu$m 
silicate bump. Only in the last decade, in the wake of the observations towards the 
Galactic Centre (Lutz et al. 1996), evidence has grown that the extinction curve can 
also undergo a flattening longward of $\sim$4~$\mu$m, without giving rise to a clear 
minimum before the silicate feature (e.g. Indebetouw et al. 2005; Nishiyama et 
al. 2009, and references therein). Since we are applying extinction to a compact 
nuclear source, this flat variant represents another reasonable choice. Moreover all 
the \textit{anomalous} sources (regardless of their absorption features) have a very 
steep \textit{L}-band continuum, suggesting that the AGN component is actually reddened and 
that some change in the extinction pattern may occur at longer wavelengths. \\ 
In order to test this scenario, we have performed our AGN/SB decomposition 
once again, this time assuming the analytical form of Chiar \& Tielens (2006) 
for the wavelength dependence of the optical depth. The intrinsic slope of the 
AGN template remains $\Gamma_\mathit{agn}$=1.5. The alternate extinction law is 
really quasi-grey in our working range, since $\tau_\mathit{red}/\tau_\mathit{blue}\simeq$0.90 
(and $\tau_\mathit{red}/\tau_6\simeq$0.96). This entails a degeneracy between 
the AGN flux amplitude and its optical depth, in spite of which we are able 
to reach important conclusions. The overall results can be summarized in terms 
of subclasses. \textit{(a)} A flat extinction can be safely ruled out for eight 
sources, whose observed continuum is too steep to be reproduced within the new 
framework. This group encompasses all the objects with $\alpha_\mathit{bol}>0.25$ 
and $\tau_6>1$ according to the previous estimate; the only exception is 
IRAS~15250+3609. \textit{(b)} The objects whose classification is unchanged 
reach up to 30, by combining 19 out of 21 pure starbursts (save IRAS~02411$+$0353 
and IRAS~16487+5447, that may contain a faint AGN component) and 11 out of 14 
composite sources hosting a non-extinguished AGN. \textit{(c)} The remaining 
three, along with IRAS~07598+6508, IRAS~19254$-$7245, the above-mentioned 
IRAS~16487+5447 and other four mildly reddened sources, have a non-degenerate 
fitting output. \textit{(d)} Finally, there are 23 objects (including the 
\textit{anomalous} ones) that can be fitted with the modified extinction but 
are affected by the normalization/absorption degeneracy. It has been possible 
to remove such degeneracy in 20 cases, by pegging the value of the optical 
depth after our previous \textit{L}-band studies. \\
\begin{figure}
\includegraphics[width=8.5cm]{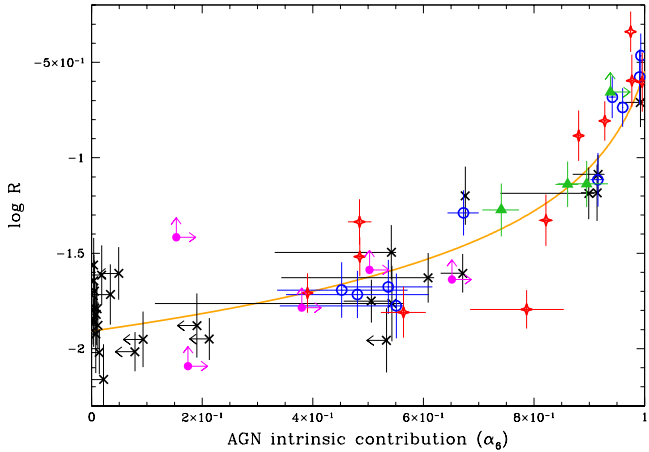}
\caption{$R$--$\alpha_6$ plot obtained by assuming a quasi-grey extinction law for 
the AGN component. The new positions of the sources are compared with those obtained 
with a power-law extinction: the \textit{orange solid line} is indeed the best fit of the 
$R$--$\alpha_6$ relation from Eq.(\ref{ra}). Different colours and symbols are used 
according to the following: \textit{black crosses} for sources confirmed to host a 
non-extinguished AGN component, if any; \textit{blue rings} for sources with a definite 
fit in the new scenario; \textit{red diamonds} for sources whose normalization has 
been fixed through their \textit{L}-band spectra; \textit{green triangles} for sources in which 
the removal of the normalization/absorption degeneracy is tentative, due to the lower 
quality of the \textit{L}-band spectra; \textit{magenta points} for sources appearing as pure 
starbursts in the \textit{L}-band.} 
\label{ext}
\end{figure}
The consequence of a quasi-grey extinction have been tested by computing 
the new values of $R$ and $\alpha_6$ and checking their re-arrangement with 
respect to the former best fit of the $R$--$\alpha_6$ relation. The outcome 
is shown in Fig.\ref{ext} and implies two interesting results: 1) a flat extinction 
law seems suitable to describe the dusty environment surrounding the AGN component 
in many ULIRGs; 2) in a statistical sense, the assumption of a quasi-grey extinction 
does not alter our global results, once the optical depth is reliably estimated 
through additional constraints. In conclusion, the large size and the spectral 
variety of our sample suggest that the AGN component inside a ULIRG can not be 
described in terms of a universal 5--8~$\mu$m extinction pattern: both a power-law 
and a soft wavelength dependence of the optical depth appear to be involved. 
In addition to a different dust amount and composition along the line of sight, 
this is possibly due also to geometrical effects, such as the orientation of the 
obscuring torus itself and/or the clumpiness of the circumnuclear star-forming 
cores. It is however important to keep in mind the limitations of our approach: 
our description does not take into account the possible effects of radiative 
transfer, that can be essential even at 5--8~$\mu$m. In particular, the deviations 
from a power-law extinction could be related to a lesser accuracy of the screen 
approximation, which is instead correct for the Galactic Centre. This issue 
represents a critical challenge for any future study of the dust properties 
and near- and mid-IR extinction in AGN and ULIRGs. 

\subsection{Trend with bolometric luminosity}

\begin{figure}
\includegraphics[width=8.5cm]{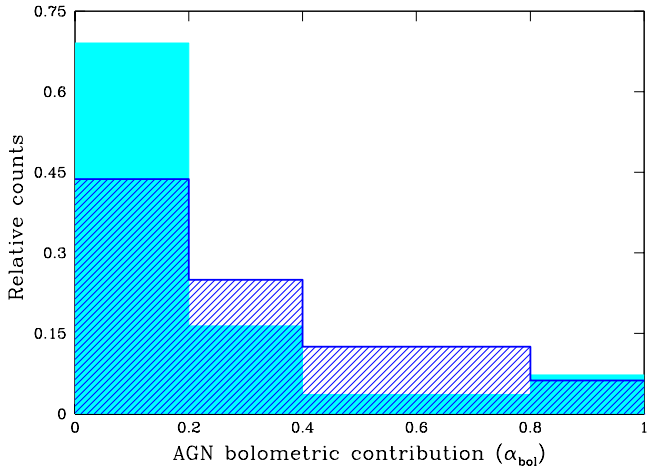}
\caption{Relative number counts within the different bins of AGN contribution. The two 
distributions correspond to the populations of sources with $\log (L_\mathit{IR}/L_\odot)<12.3$ 
(\textit{cyan}) and $\log (L_\mathit{IR}/L_\odot)>12.3$ (\textit{shaded blue}).} 
\label{th}
\end{figure}

Once the relative AGN/SB fraction of ULIRG luminosity has been estimated, 
one of the main issues to investigate is the possibility of a larger AGN 
contribution at higher luminosities. Interestingly, no clear trend in the 
detection of [Ne~\textsc{v}] $\lambda$14.32 as a function of the total IR 
luminosity has been discovered by Farrah et al. (2007). This is in contrast 
with several studies based on optical spectroscopy, which instead suggest 
the existence of a growing trend. This claim finds the first convincing 
confirmation with the \textit{histograms} of Veilleux et al. (1999a), 
indicating that AGN-like systems (i.e. those classified as Seyferts, 
without considering the possible integration of LINERs) represent only 
a small percentage among IR galaxies below $L_\mathit{IR}<10^{11} L_\odot$, 
but increase to become the most populated subset in the luminosity range 
$10^{12.3} L_\odot < L_\mathit{IR} < 10^{12.8} L_\odot$. Moreover virtually 
all the hyperluminous IR sources (HLIRGs, $L_\mathit{IR}>10^{13} L_\odot$) 
seem to harbour a powerful AGN, at least up to $z \la 1$ (Rowan-Robinson 2000). 
The key point is that such a result still has to be read in terms of AGN 
detectability, and not necessarily of larger AGN contribution. In a skeptical 
perspective even this detectability trend could be contrived by selection 
effects: as we have already pointed out, a considerable amount of sources 
that are optically classified as H~\textsc{ii} regions actually harbour 
obscured but non-negligible AGN components. Furthermore, a general consensus 
has grown in the last years about the sharp decrease in the fraction of absorbed 
AGN with X-ray luminosity (Ueda et al. 2003; Hasinger 2008). At mid-IR 
wavelengths a clear AGN preponderance at high luminosity is found by Tran et al. 
(2001), with starbursts and AGN dominating the energy output respectively at the 
lower and higher ends of the ULIRG range. It is then worthwhile checking how our 
results fit into this question, even though our coverage of bolometric luminosity 
is limited. \\
Assuming the standard divide of $\log(L_\mathit{IR}/L_\odot)=12.3$, the population 
of the two subsets consisting of the sources whose luminosity is below and above 
this threshold is 55 and 16, respectively. We recall that simply by solving the 
ambiguity concerning LINERs we are able to detect a much larger number of AGN (50 
out of 71 ULIRGs) with respect to optical studies. The positive detections involve 
$\sim$2/3 of the \textit{faint} sources and $\sim$3/4 of the \textit{bright} 
ones. Such fractions are too similar to hint at a clear increase. Nevertheless 
by computing the AGN contribution $\alpha_\mathit{agn}$ to both bins of luminosity 
we obtain two well-separated values: 
\begin{displaymath}
\alpha_\mathit{agn}=\frac{\sum \alpha_i L_i}{\sum L_i}
=\left \lbrace \begin{array}{ll}
18.2^{+0.6}_{-1.0} & (\log (L_\mathit{IR}/L_\odot)<12.3) \\
 & \\
30.9^{+1.4}_{-1.3} & (\log (L_\mathit{IR}/L_\odot)>12.3)
\end{array} \right .
\end{displaymath}
where $\alpha$ and $L$ stand for the previous $\alpha_\mathit{bol}$ and 
$L_\mathit{IR}$. 
This is robust evidence in favour of the growing trend with bolometric 
luminosity of the AGN content inside ULIRGs\footnote{There is indeed 
a difference in the cosmological parameters between this work and that 
of Kim \& Sanders (1998), on which the mentioned investigations of the 
1~Jy sample at optical wavelengths are based. This slightly affects the 
estimate of the bolometric luminosities, with an upward shift of 
$\sim$0.1~dex at the highest luminosity distances: about ten sources 
enter the higher luminosity bin. We have taken into account this effect, 
and found that the trend is both qualitatively and quantitatively confirmed.}. 
An illustration of this effect, that will be addressed in detail in a 
forthcoming paper (Nardini et al., in preparation) by enlarging the statistics 
at the highest luminosities (see also Imanishi 2009), is provided in 
Fig.\ref{th}. \\
In conclusion, considering only the 63 sources from the 1 Jy sample, AGN are 
responsible for $23.1^{+0.7}_{-0.9}$ per cent of the ULIRG luminosity in the 
local Universe. We underline that this estimate is not biased due to selection 
effects. Concerning the seven missing sources which undermine the sample completeness, 
by simply assuming the average properties of their optical class for the five 
type 2 Seyferts (i.e. $\alpha_\mathit{bol} \simeq$0.29) and a SB-dominated nature 
for the two unclassified objects we are able to confirm the value provided above. 
Only in the extreme case of null AGN significance for all the seven sources the 
global contribution of accretion processes to IR activity should be reduced by 
$\sim$2 per cent. The same amplitude in the opposite direction is predicted after 
correcting for a possible bias towards star formation, connected to the 
\textit{cold} 60~$\mu$m selection. Not even the inclusion of the eight 
southern ULIRGs in the computation affects our results, since they form a complete 
and fully representative subset. \\
The recent and comprehensive work of Veilleux et al. (2009) investigates with great 
detail the connection between local ULIRGs and PG quasars; the contribution of nuclear 
activity to the bolometric luminosity of ULIRGs is obtained by means of six independent 
methods based on the \textit{Spitzer}-IRS spectra. Fine-structure line ratios, aromatic 
features and continuum colours have been employed: a comparison among the six different estimates 
provided for the single sources gives a good idea of the uncertainties involved, that sometimes 
can be pretty large. The ensemble average hints at a slightly higher AGN contribution 
($\sim$35--40 per cent) with respect to the findings of our study and the previous 
literature. This could be due to a possible bias in Veilleux et al. (2009), related to the 
choice of PG quasars as \textit{zero points} for the nuclear activity within ULIRGs. 
However, by simply restricting to the 48 common sources, our estimate increases to 
$\sim$30 per cent as well, and the results are definitely in good agreement.


\section{Conclusions}

In this paper we have presented a 5--8~$\mu$m spectral study of a large 
and representative sample of local ULIRGs, based on \textit{Spitzer}-IRS 
observations. We have explored the role of black hole accretion and vigorous 
star formation as the power source of the extreme IR activity, assessing 
the gravitational and stellar origin of ULIRG emission by modelling their 
5--8~$\mu$m spectra through a couple of AGN and SB templates. In this 
wavelength range, in fact, the spectral properties of active galactic 
nuclei and starbursts are widely different, and only moderate dispersion 
is found within each class. It is then possible to accomplish a sharp 
characterization of both the AGN and SB components, and disentangle their 
contribution to the observed emission. Our method has proven to be successful 
in unveiling intrinsically faint or heavily obscured AGN components, leading 
to a total of 50 convincing detections out of the 71 sources in our sample. 
Such a large AGN detection rate can be achieved only by resorting to the 
most effective multiwavelength diagnostics. As a consequence, AGN turn out 
to be very common among ULIRGs in the local Universe, even if they are usually 
not significant as contributors to the global energetics of the host galaxy 
when compared to the starburst. \\
A simple analytical model, in fact, allows us also to obtain a reliable  
estimate of the AGN/SB contribution to the overall energy output of each 
source. For the sake of a gross classification we can define the following 
\textit{one third rule}: assuming a significance threshold of 1/3 in 
the AGN/SB bolometric contribution, i.e. $\alpha_\mathit{bol}=0.25$, we find that 
one third of ULIRGs harbour a sizable AGN, one third harbour a negligible AGN, 
and one third harbour no AGN at all. Star formation is then confirmed as 
the dominant energy source underlying the ULIRG activity. In more quantitative terms, 
AGN account for $\sim$23 per cent of the global ULIRG emission, approaching and 
possibly exceeding $\sim$30 per cent at higher luminosities. This 
increasing trend is clearly brought out: the average AGN \textit{weight} 
rises from 18 to 31 per cent above the luminosity threshold of 
$L_\mathit{IR}=10^{12.3} L_\odot$. \\
Thanks to the large size of our sample we have also been able to test 
the properties of the obscuring material, in particular of the extinction 
law applicable to the AGN-related hot dust continuum. Observations reveal 
a lack of correlation between the continuum reddening and the presence of 
deep absorption features, suggesting that the extinction of the AGN component 
in a ULIRG environment is not universal. Both a power-law and a quasi-grey 
behaviour of the optical depth as a function of wavelength are necessary to 
account for the emission of different objects and seem to be involved among 
ULIRGs. Large differences in the chemical composition and geometrical structure 
of the dust are supported by the spread in the intensity (and sometimes also 
the shape) of the most frequent absorption profiles. Although our method is 
not suitable to fully investigate these aspects, it is extremely powerful for 
global classification and can be applied whenever spectra of sufficient quality 
are available. Moreover it can be turned into a photometric method to study 
fainter sources, using diagnostics such as the bolometric correction and the 
continuum slope, which can be measured efficiently with the upcoming IR facilities.

\section*{Acknowledgments}

We are grateful to the anonymous referee for the constructive comments and 
suggestions that improved our work. This research has made use of the NASA/IPAC 
Extragalactic Database (NED) which is operated by the Jet Propulsion Laboratory, 
California Institute of Technology, under contract with the National Aeronautics 
and Space Administration. We acknowledge financial support from PRIN-MIUR 
2006025203 grant, and from ASI-INAF I/088/06/0 and ASI-INAF I/016/07/0 contracts. 



\appendix
\section{Additional notes and spectral fits}

With the exception of IRAS~08559+1053, that was observed in the early stage 
of the mission (before the start of the regular scientific operations) and is analyzed 
for the first time in the present work, the low- and/or high-resolution 
\textit{Spitzer}-IRS spectra of the sources in our sample have already been shown 
and widely discussed by Armus et al. (2007), Farrah et al. (2007), Imanishi et al. (2007), 
and Veilleux et al. (2009). In the following we add some additional comments about a few 
interesting objects, and show the full results of our spectral decomposition. 

\subsection{The nature of ARP~220} 
In spite of being the nearest (at $\sim$80~Mpc) and best-studied ULIRG, ARP~220 
is still puzzling and controversial with respect to its energy source. The multiwavelength 
SED of this interacting system has been often employed as a template 
to constrain the emission of star-forming galaxies at high redshift, but the pure 
SB scenario is actually challenged by many independent pieces of observational 
evidence, and the alleged action of an extremely obscured and elusive black hole 
in the western nucleus is widely debated. Such a possibility is strengthened by 
recent interferometric observations, outlining a compact and massive ring of hot 
dust that straightly recalls the toroidal structures around AGN (Downes \& Eckart 2007, 
and references therein for additional hints to the presence of an AGN inside ARP~220). 
It is somewhat arbitrary to assess the intrinsic luminosity of this core; the western 
nucleus is anyway the brightest one by a factor $\sim$3 (Soifer et al. 1999) and 
the AGN component, if confirmed, would be absolutely relevant. Our diagnostic method 
discloses a contribution to the bolometric luminosity of the whole system arising from 
a highly-obscured AGN, that reaches up to 20 per cent. This value should be read with some 
caution, also due to the peculiar properties of the aromatic emission in ARP~220 
(e.g. the suppression of the 6.2~$\mu$m feature with respect to the 7.7~$\mu$m one) 
that may alter the continuum decomposition based on AGN/SB templates; yet it 
represents a stirring indication of magnitude. The matter is fully open, and further 
incentive is provided by Sakamoto et al. (2008), whose precise discussion about the 
nature of the compact luminous source located in the western nucleus, both in terms 
of an AGN and of a collection of overlapping super star clusters, makes the latter 
explanation appear as slightly strained. 

\subsection{The iron K$\balpha$ line in IRAS~04103$-$2838} 
IRAS~04103$-$2838 is optically classified as a LINER and is a \textit{warm} ULIRG according 
to its IRAS colours ($f_{25}/f_{60}=0.30$). We are able to unveil an AGN component at 
5--8~$\mu$m, whose contribution to the bolometric luminosity is estimated as $\sim$6 
per cent. The presence of an AGN is confirmed by the detection of a broad K$\alpha$ line at 
$\sim$6.4~keV, during a recent \textit{XMM-Newton} observation (Teng et al. 2008). 
Due to the large equivalent width ($\sim$1.65~keV) of this iron feature the AGN is 
supposed to be Compton-thick, anyway it does not seem to dominate the total energy 
output of the source.

\subsection{The 7.65~$\bmu$m fine-structure line}
IRAS~17179+5444 exhibits a prominent emission at $\sim$7.65~$\mu$m, blended with the adjacent 
PAH feature, that we interpret as the unresolved [Ne~\textsc{vi}] fine-structure line. 
Due to its ionization energy of 126~eV this feature is connected to the hardest nuclear 
activity, and indeed it is convincingly observed also in other three Seyfert galaxies 
(IRAS~05189$-$2524, IRAS~12072$-$0444 and IRAS~13454$-$2956) plus IRAS~04103$-$2838. 
Fluxes and equivalent widths are listed in Tab.\ref{t4}, and are in good agreement 
with the values provided by Veilleux et al. (2009) for their positive detections. We note 
that our identification of this line also in IRAS~04103$-$2838 represents the first detection 
involving a non-Seyfert ULIRG. In IRAS~17179+5444 the strength of [Ne~\textsc{vi}] is anomalously 
high, as evinced through both a visual inspection and a comparison with the [Ne~\textsc{v}] lines 
at longer wavelengths (Farrah et al. 2007). This is a very interesting case, since after a separate 
check of the single spectra from each nod position the line is confirmed to be genuine. 

\begin{table*}
\begin{center}
\caption{Flux and equivalent width (with respect to the AGN component) of the 7.65~$\mu$m 
[Ne~\textsc{vi}] fine-structure line. The uncertainties on both quantities range from $\sim$5 
to 30 per cent.}
\label{t4}
\begin{tabular}{lcc}
\hline
Source & [Ne~\textsc{vi}] Flux & $EW$ \\
(IRAS) & (ergs s$^{-1}$ cm$^{-2}$) & ($\mu$m) \\ 
\hline
04103$-$2838 & $1.0\times10^{-13}$ & 0.024 \\
05189$-$2524 & $1.1\times10^{-13}$ & 0.008 \\
12072$-$0444 & $2.3\times10^{-14}$ & 0.011 \\
13454$-$2956 & $1.5\times10^{-14}$ & 0.040 \\
17179+5444 & $3.8\times10^{-14}$ & 0.057 \\
\hline
\end{tabular}
\end{center}
\end{table*}

\subsection{The case study of NGC~6240} 
As mentioned, many X-ray observations suggest the possible presence in NGC~6240 
of a pair of supermassive black holes hidden behind large column densities 
of screening material. A detailed IR study pivoted on the \textit{Spitzer}-IRS 
spectrum and integrated with photometric data from near- to far-IR is presented 
by Armus et al. (2006). As a result of the SED modelling a hot dust component at 
$\sim$700~K is found, accounting for $\sim$3.5 per cent of the bolometric luminosity 
and pointing to the presence of a buried AGN. The AGN contribution is again 
estimated as $\sim$3--5 per cent through high-ionization lines, but can raise up to 
$\sim$20--24 per cent if correcting the [Ne~\textsc{v}]/[Ne~\textsc{ii}] ratio for 
the extinction derived from the hard X-ray data. Our estimate of 
$\alpha_\mathit{bol}$ (4.5--11.5 per cent) is then fully reliable, in spite 
of the great simplicity of our method as opposed to a multicomponent 
fit of the whole IR SED, which is anyway necessary to fully characterize 
an individual source. 

\subsection{Spectral fits}
The following panels illustrate the results of our AGN/SB decomposition 
at 5--8~$\mu$m. Along with the data points (\textit{green filled circles}) 
and the best fit model (\textit{black solid line}), both contributions are 
plotted: the SB component (\textit{red dot-dashed line}) and the observed 
AGN continuum (\textit{blue long-dashed line}). 

\begin{figure*}
\includegraphics[width=\linewidth]{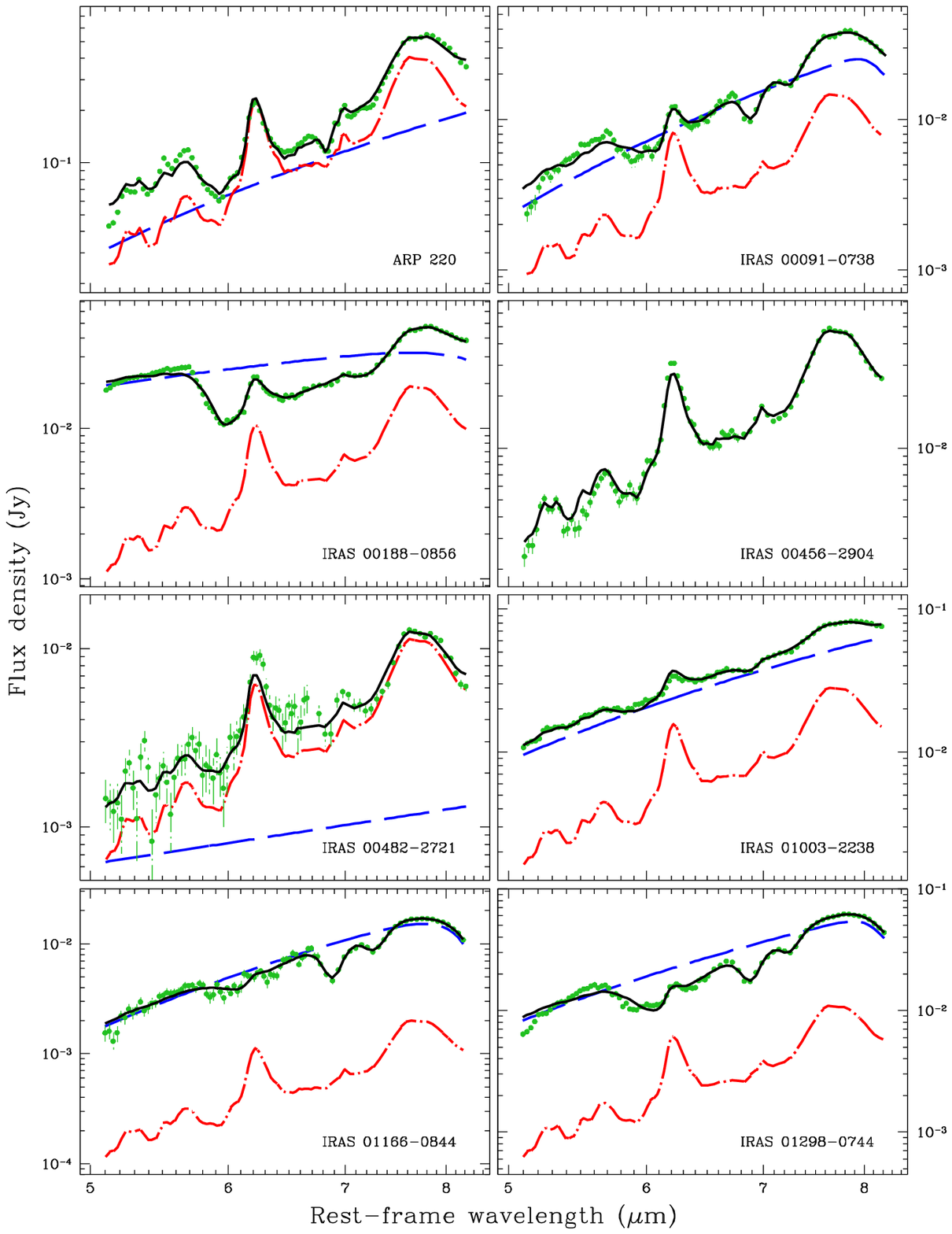}
\caption{Best fits and spectral components.} 
\end{figure*}
\begin{figure*}
\includegraphics[width=\linewidth]{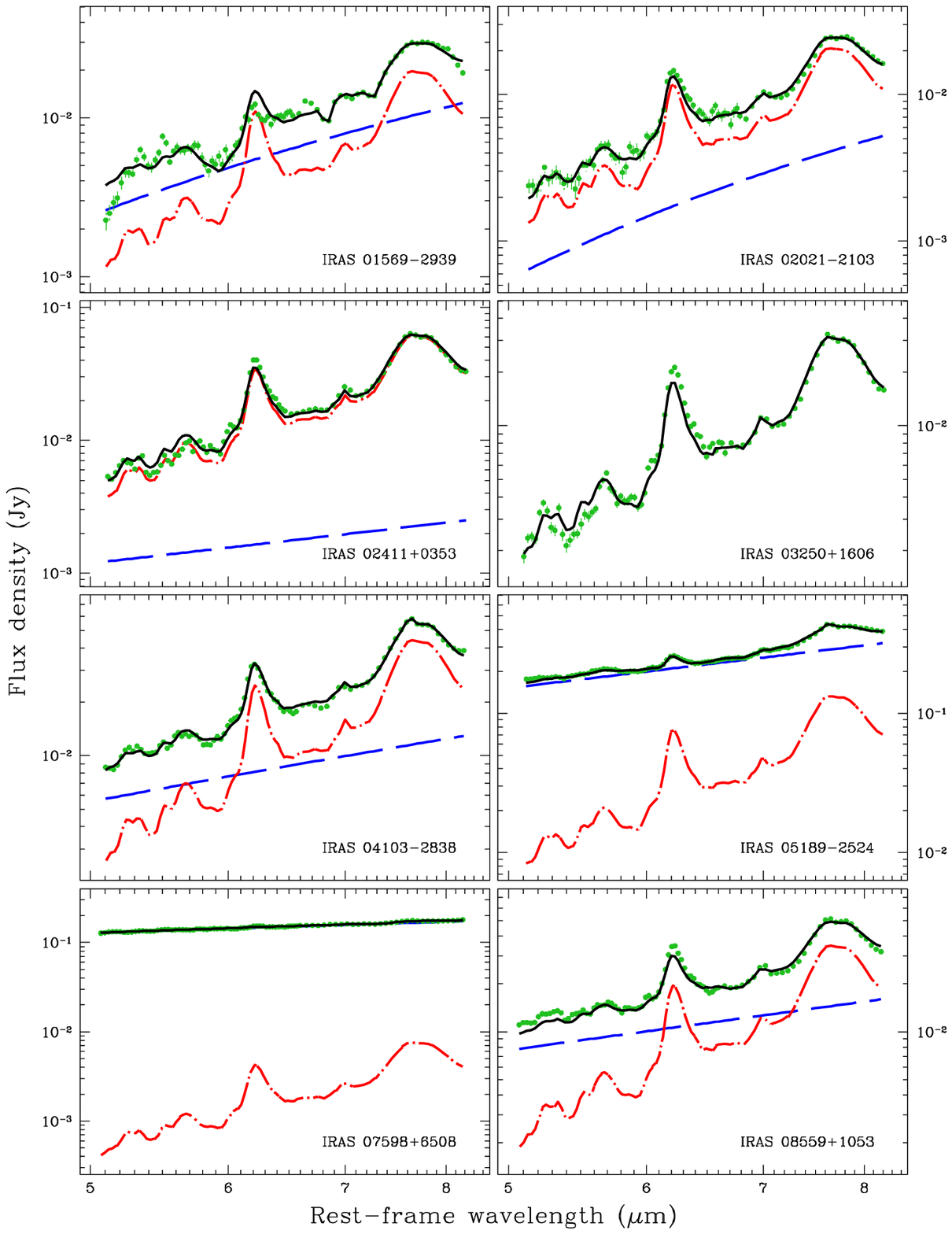}
\contcaption{} 
\end{figure*}
\begin{figure*}
\includegraphics[width=\linewidth]{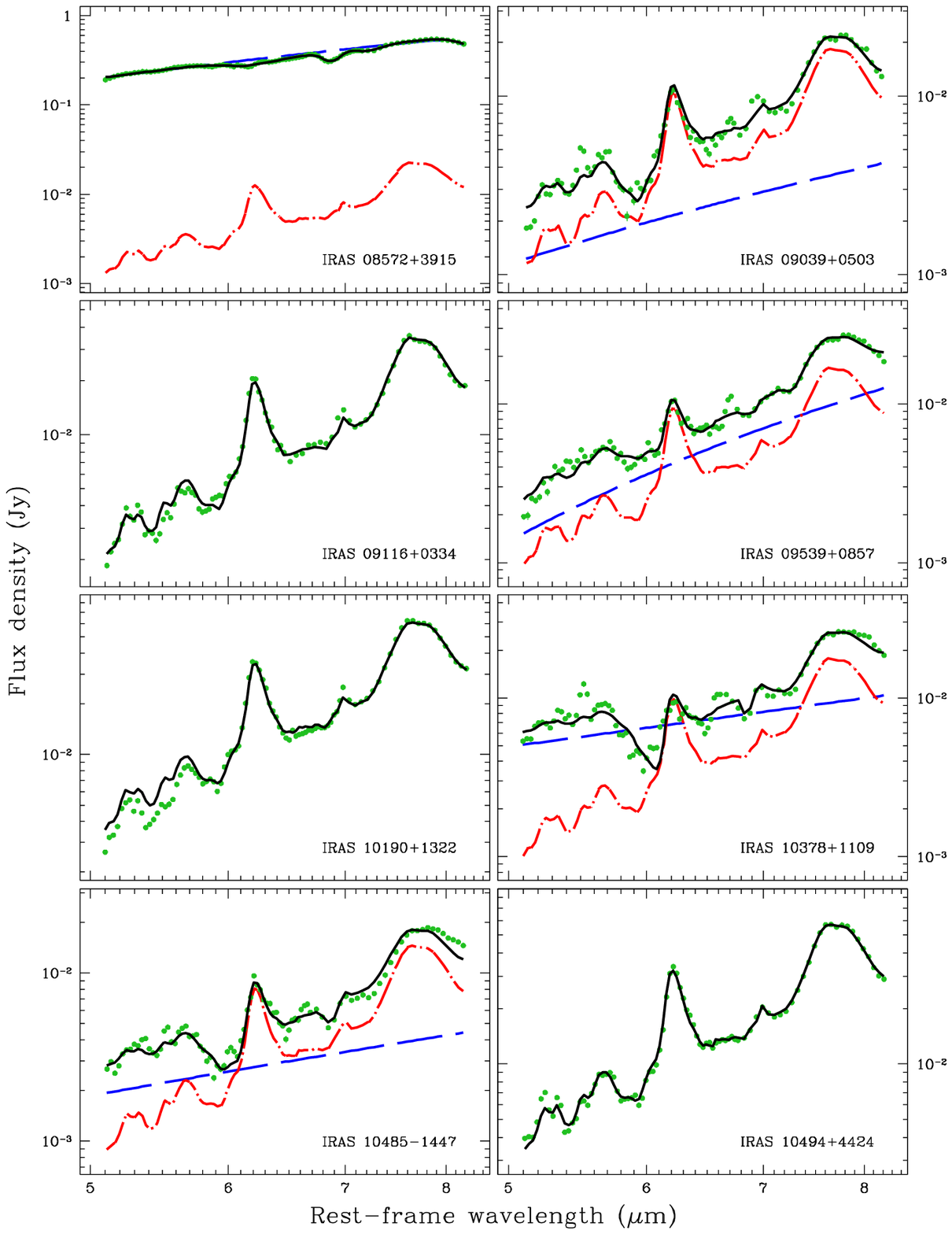}
\contcaption{} 
\end{figure*}
\begin{figure*}
\includegraphics[width=\linewidth]{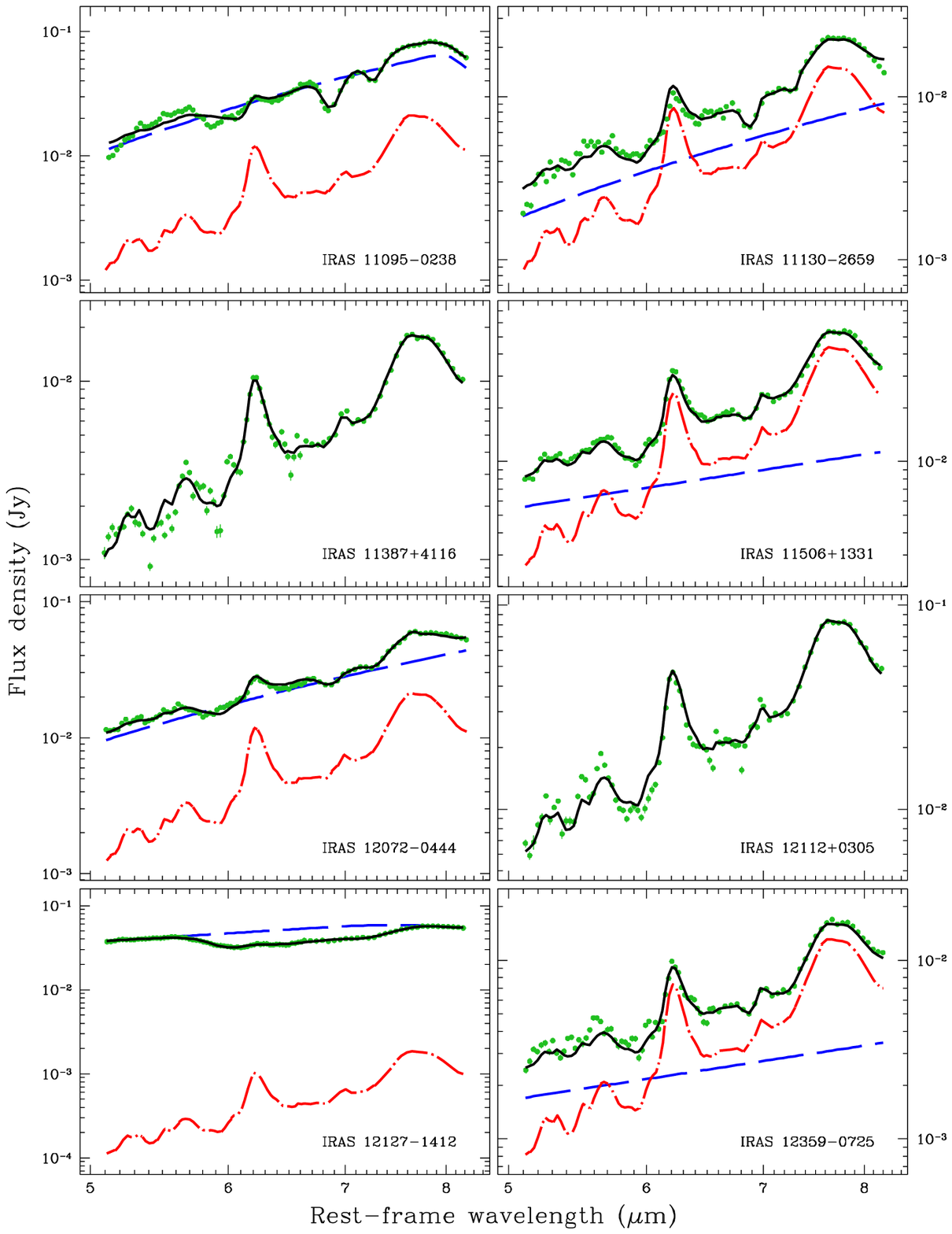}
\contcaption{} 
\end{figure*}
\begin{figure*}
\includegraphics[width=\linewidth]{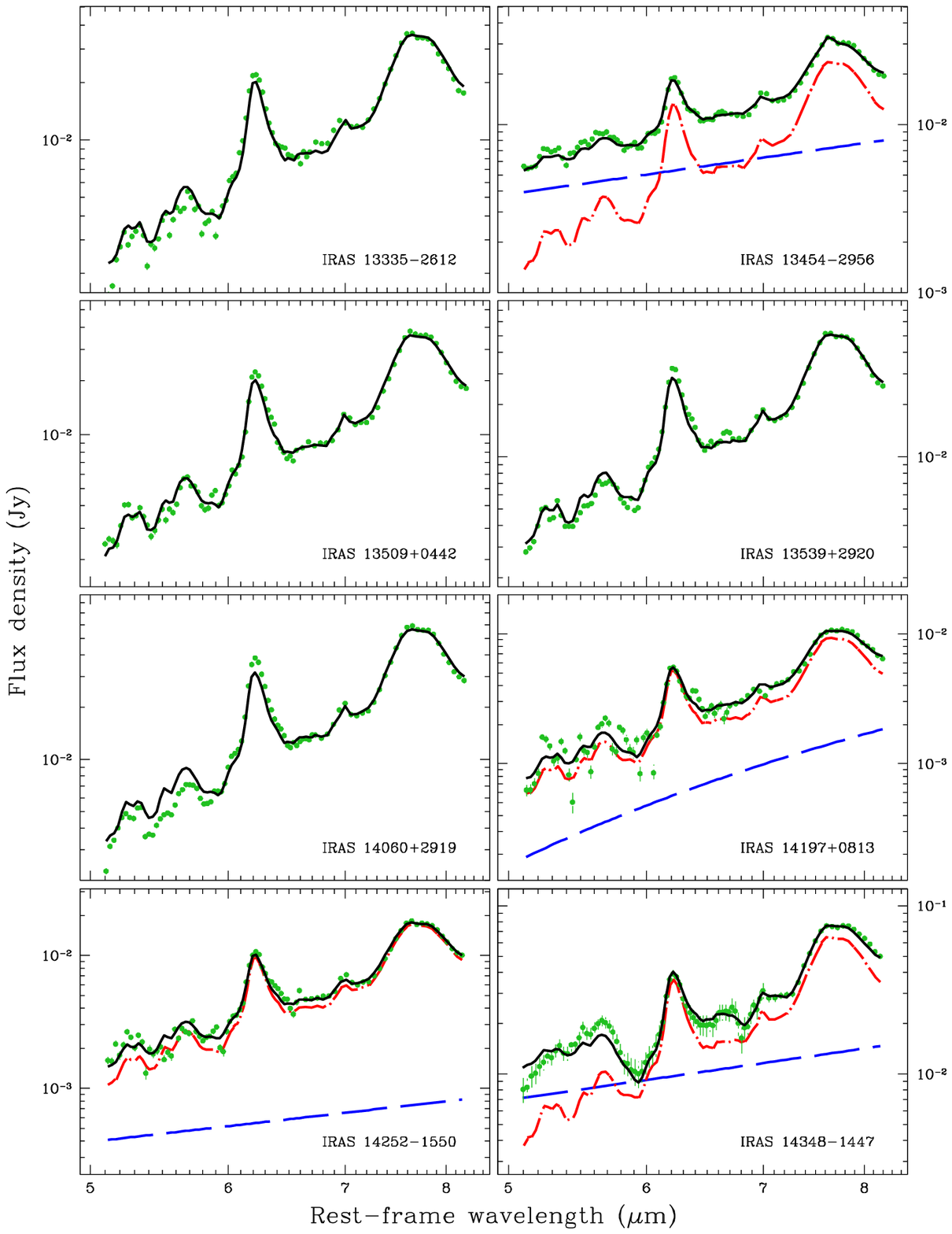}
\contcaption{} 
\end{figure*}
\begin{figure*}
\includegraphics[width=\linewidth]{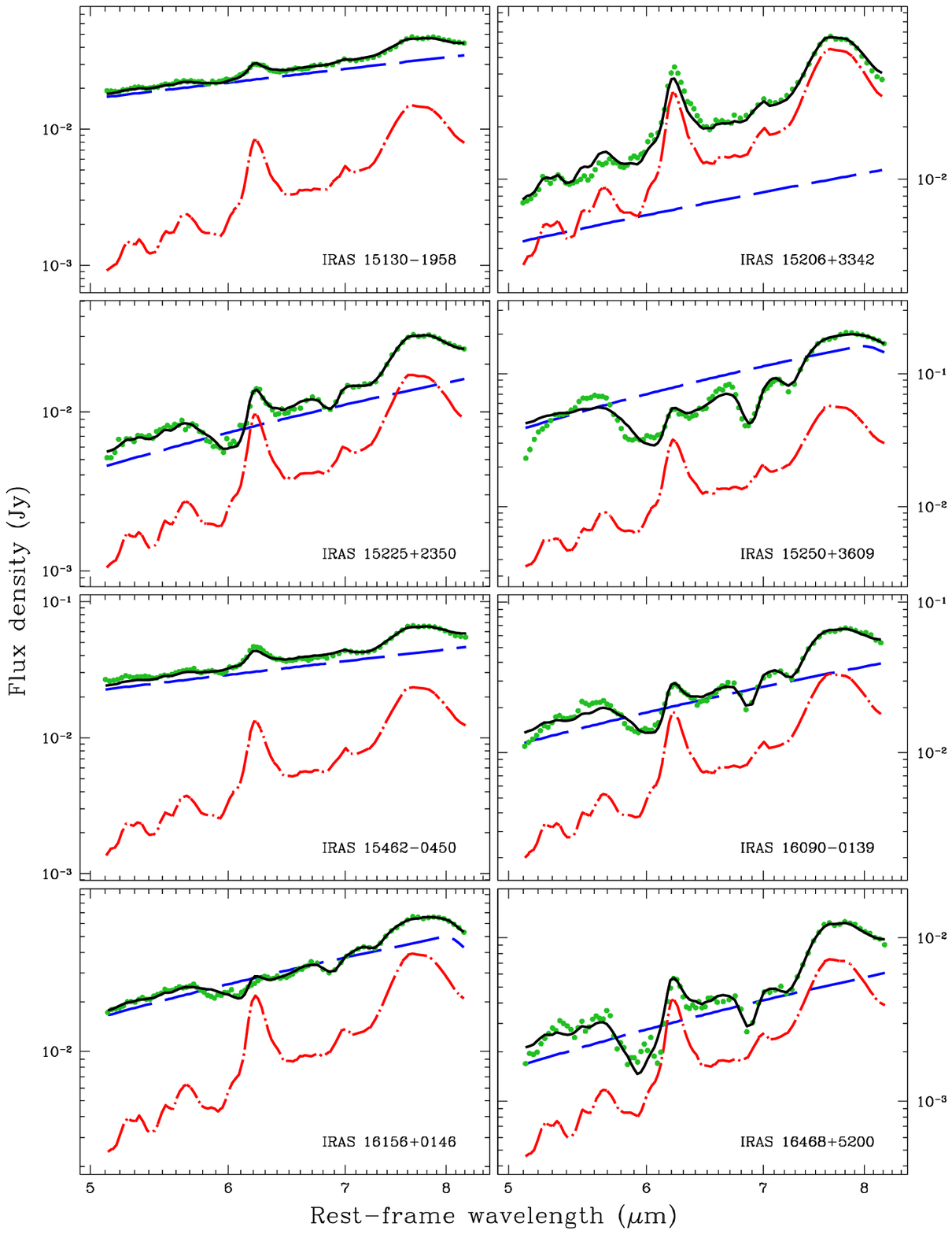}
\contcaption{} 
\end{figure*}
\begin{figure*}
\includegraphics[width=\linewidth]{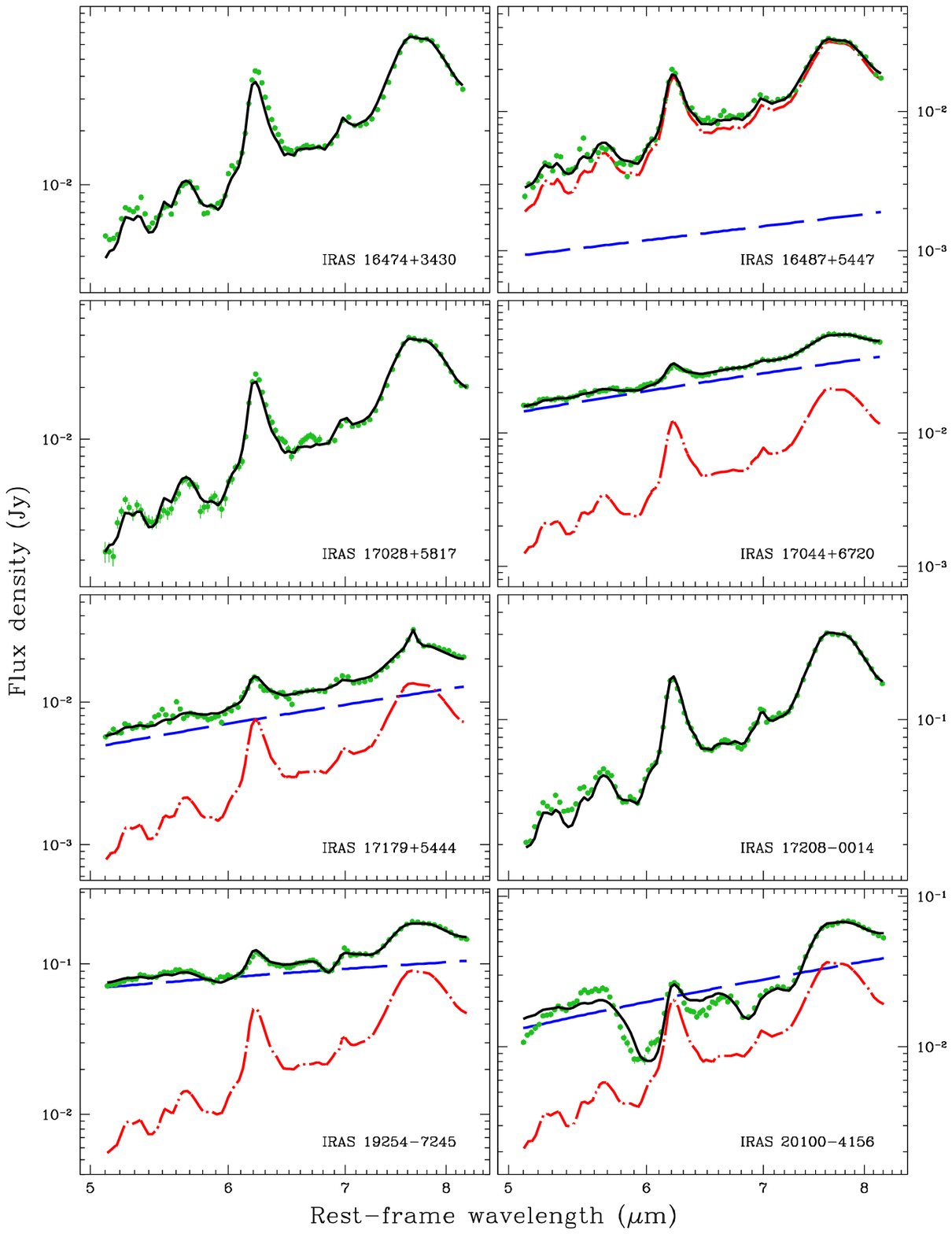}
\contcaption{} 
\end{figure*}
\begin{figure*}
\includegraphics[width=\linewidth]{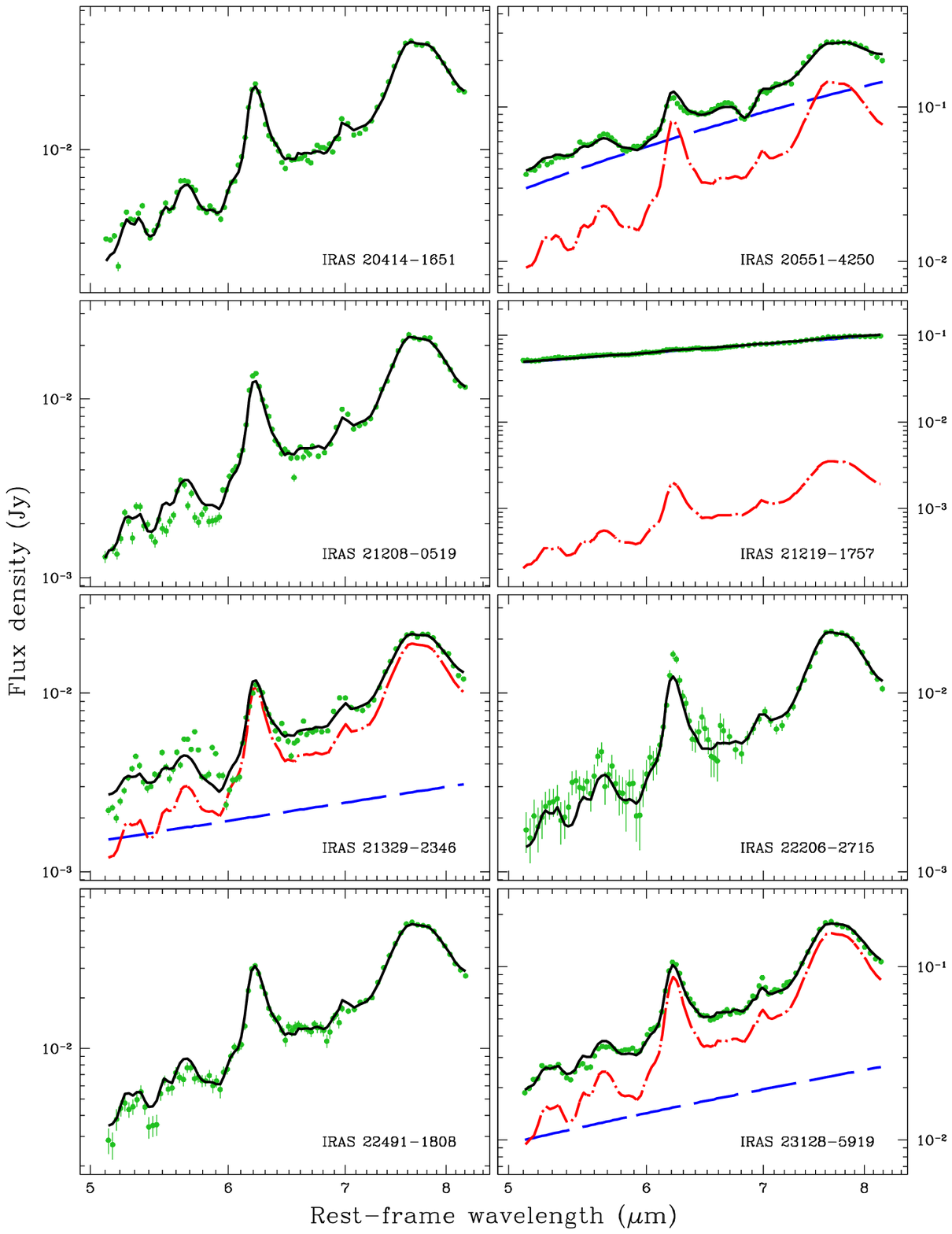}
\contcaption{} 
\end{figure*}
\begin{figure*}
\includegraphics[width=\linewidth]{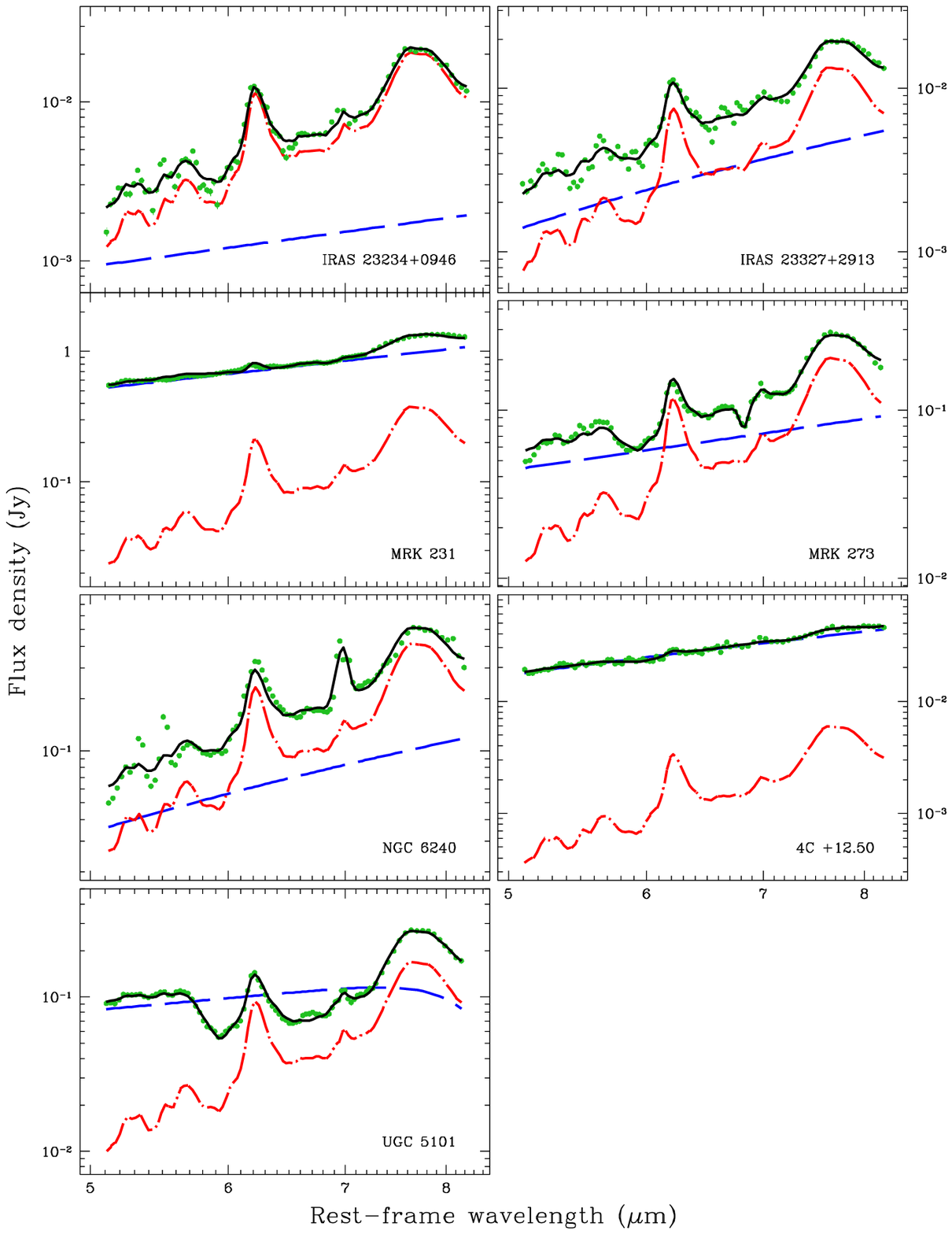}
\contcaption{} 
\end{figure*}

\label{lastpage}

\end{document}